\documentclass[aps, prb, superscriptaddress, reprint, twocolumn]{revtex4-2}

\usepackage{changes}
\usepackage{amsmath}
\usepackage{amsfonts}
\usepackage{amssymb}
\usepackage{gensymb}
\usepackage{graphicx}
\usepackage[utf8]{inputenc}
\usepackage[colorlinks=true, linkcolor=black, urlcolor=blue, citecolor=black, anchorcolor=black]{hyperref}
\usepackage{enumitem}

\setlist{nosep}

\newif\iffigures
\figuresfalse 
\figurestrue

\begin{document}
\title{Trapping Interlayer Excitons in van der Waals Heterostructures by Potential Arrays}

\author{Darien J. Morrow}
\affiliation{Center for Nanoscale Materials, Argonne National Laboratory, Lemont, Illinois 60439, United States}

\author{Xuedan Ma}
\email{xuedan.ma@anl.gov}
\affiliation{Center for Nanoscale Materials, Argonne National Laboratory, Lemont, Illinois 60439, United States}
\affiliation
{Consortium for Advanced Science and Engineering, University of Chicago, Chicago, Illinois 60637, United States}
\affiliation
{Northwestern-Argonne Institute of Science and Engineering, 2205 Tech Drive, Evanston, IL 60208, USA}

\date{\today}

\begin{abstract}
	Transition metal dichalcogenide heterostructures can host interlayer excitons (IXs), which consist of electrons and holes spatially separated in different layers. IXs possess permanent dipoles and have proven to offer a wealth of novel physics.
 	We develop a discrete, random-walk model which includes annihilation and repulsion interactions among IXs. Using this model, we simulate the trapping of IXs in traps of different depths, densities, and shapes. Our results show that dipole-dipole interactions play an important role in regulating IX trapping. The effects of dipole interactions can be mitigated with small, deep traps which are realizable with atmoic defects and moir\'e potentials.
\end{abstract}

\maketitle

\section{Introduction}

An interlayer exciton (IX) is composed of an electron and hole that are spatially separated in different layers but bound by Coulomb interactions. Due to the spatial separation of the constituent electrons and holes, IXs can have lifetimes that are orders of magnitude longer than those of intralayer excitons.\cite{Rivera_Xu_2015} Furthermore, IXs possess permanent, out-of-plane electric dipole moments, rendering their binding energies effectively tunable using external electric fields.\cite{Ciarrocchi_Kis_2018, Karni_Heinz_2019} 
These characteristics may allow IXs to be cooled below the temperature of quantum degeneracy, leading to the realization of quantum Bose gases.\cite{Butov_Gossard_2001, Wang_Mak_2019} 
Their widely tunable excitonic response is also promising for efficient light emitting and modulation applications.\cite{Lukin_Zoller_2001, Paik_Deng_2019} 

While indirect excitons in conventional III-V semiconductor heterostructures have been extensively studied,\cite{Combescot_Dubin_2017, High_Gossard_2012_B} those in two-dimensional transition metal dichalcogenide (TMD) heterostructures provide alternative opportunities for exploring excitonic quantum gases and devices at high temperatures\cite{Fogler_Novoselov_2014} due to their considerably larger binding energies.\cite{Gillen_Maultzsch_2018, Donck_Peeters_2018} The spin-valley degree of freedom offered by IXs in TMDs may also enable unconventional strategies for data processing and transmission.\cite{Zhang_Yu_2019} 

Exploration of IXs in TMD heterostructures for these applications almost exclusively rely on their trapping and localization. In order to obtain a cold and dense exciton gas for their spontaneous condensation, a potential trap is often required to accumulate the indirect excitons towards the bottom of the trap potential.\cite{High_Gossard_2009_A, Yoshioka_Kuwata-Gonokami_2011} The transport of IXs can be modulated by their localization-delocalization transitions across lattice potentials.\cite{Remeika_Gossard_2009, Yuan_Huang_2020} Moreover, the strong repulsive dipole-dipole interactions among IXs can be engineered by confining them in potential traps created by strains, disorders or moir\'{e} potentials,\cite{Wang_Ma_2020, Shanks_Schaibley_2021, Wu_MacDonald_2018} which may provide rich opportunities for the study of quantum optical nonlinearities and creation of quantum photon sources.\cite{Imamoglu_Deutsch_1997, Peng_Ma_2020, Baek_Gerardot_2020}

While great strides have been made in trapping IXs in TMDs using strains and moir\'{e} potentials,\cite{Wang_Ma_2020, Wu_MacDonald_2018, Kremser_Finley_2020, Cho_Atwater_2021} the influence of the potential trap nature, such as the trap geometries and densities, on the related photodynamics of the IXs including their trapping and localization remain largely unexplored. 
Specifically, upon excitation of TMD heterostructures, IXs can be formed over picosecond timescales.\cite{Ovesen_Malic_2019, Ceballos_Zhao_2014} The longer lifetimes of the IXs compared to those of the intralayer excitons indicate the potentially larger diffusion lengths of the IXs and their higher probabilities to encounter potential traps before recombination occurs.\cite{Harats_Bolotin_2020_B}
The competition between the trapping and recombination of the IXs should be heavily influenced by the density and arrangement of the potential traps. Additionally, IXs may undergo trapping-detrapping dynamics depending on the experienced thermal fluctuations and dipole-dipole interactions, which are closely related to the system temperature and trap dimensions.\cite{Wang_Ma_2020, Li_Deotare_2021} These factors, combined with the intrinsic electronic structures of the TMD monolayers manifested as bright and dark excitonic states,\cite{Yuan_Huang_2017, Echeverry_Gerber_2016} constitute multiple degrees of freedom that are important to control when creating TMD devices with optimal IX trapping efficiencies.

Here, we systematically investigate the trapping and localization of IXs based on a discrete-time random walk model. By including realistic inter-exciton processes like exciton-exciton annihilation and dipole-dipole repulsion, we investigate the effects of trap dimensions and densities on IX dynamics. We show the trade-off between trap parameters like depth, radius, and density compared to externally controlled parameters like temperature and initial exciton density in achieving efficient IX trapping. We find that trap radius and dipole-dipole interaction among IXs play a dominant role in determining their trapping efficiencies. Our results evidence that myriad small traps at low temperature yield the highest trapping efficiency of IXs. These findings illustrate the influence of various parameters on IX dynamics, and  provide heuristics for the design of TMD heterostructures with optimized IX trapping properties which are targeted for quantum  nonlinearity and quantum photon source applications.

\section{Theoretical Model}

\begin{figure*}[!htbp]
	\centering
	\iffigures
	\includegraphics[trim=1.5cm 1.cm 1.5cm 2.cm, clip,width=\linewidth]{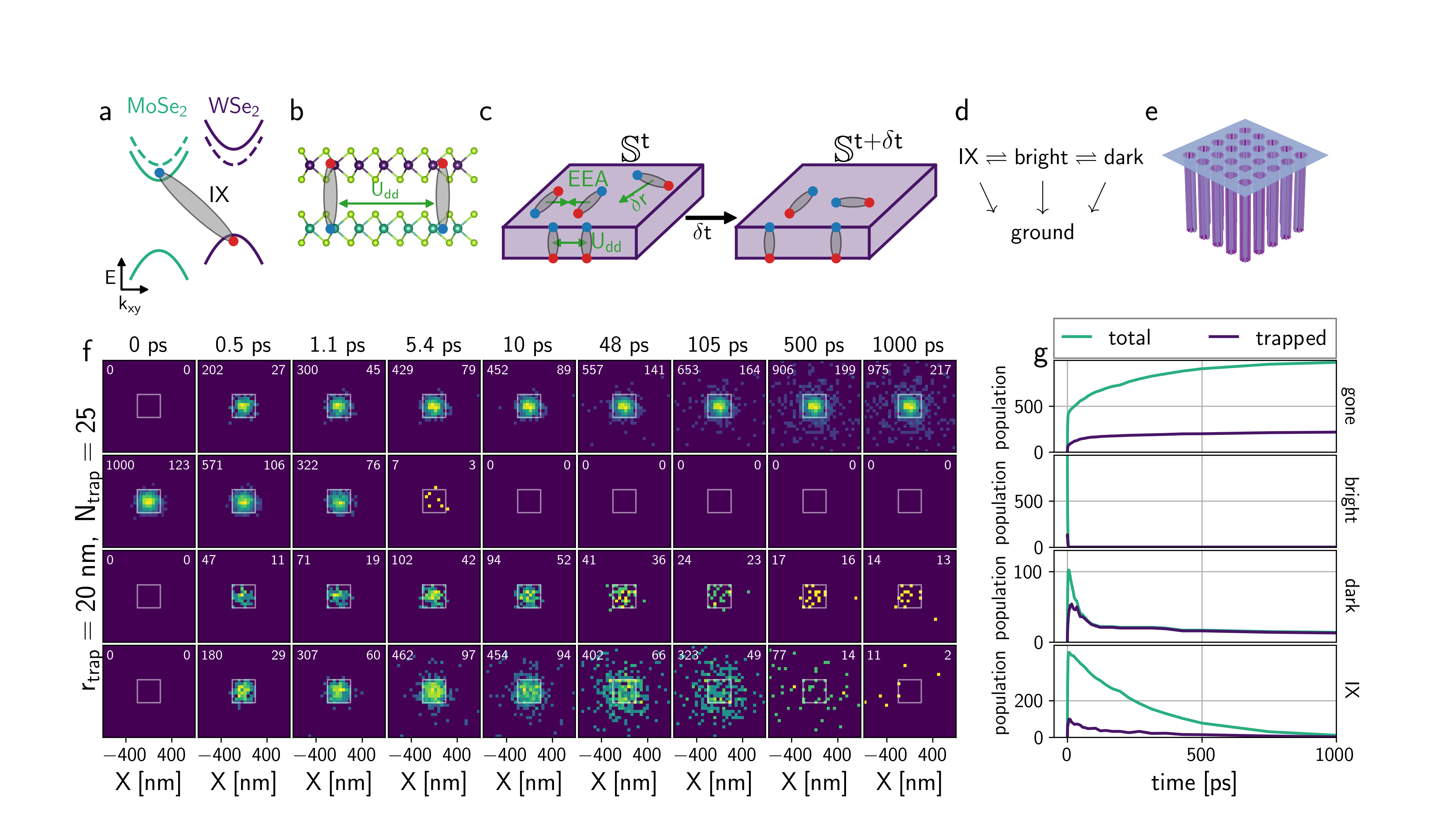}
	\fi
	\caption{
		Overview of exciton trapping simulation.
		(a, b) Sketch of a MoSe\textsubscript{2}/WSe\textsubscript{2} heterostructure with a type-II band alignment.
		The bright and dark band-edge states are represented by solid and dashed lines, respectively.
		(c) Abstraction of intra- and interlayer exciton dynamics to a single plane with exciton-exciton annihilation (EEA), diffusion, and IX dipole-dipole interactions ($U_{dd}$). 
		(d) Interconnections between IX, bright \& dark intralayer excitons, and ground states. 
		(e) Sketch of a 5 $\times$ 5 flat-bottomed potential trap array.
		(f) Evolution of excitons through space and time with conditions: $T=5\text{ K}$, $E_{\text{trap}}=50\text{ meV}$, $N_{\text{ex},0}=1000$. The potential trap array consists of $N_{\text{trap}} = 5 \times 5 = 25$  flat-bottomed traps each with a radius, $r_{\text{trap}}$, of 20 nm that are all equally separated by 50 nm. 
		The convex hull of the traps is boxed in white.
		The rows correspond to different types of excitons: decayed (gone), bright, dark, and interlayer exciton (IX). The total number of the specific type of excitons is shown in the top left corner of each frame while the number of the corresponding trapped excitons is in the top right corner. 
		(g) The total and trapped populations of each exciton type at each time point of the simulation shown in (f) on a linear-linear scale.    
	}\label{fig1:intro} 
\end{figure*}

In this work, we consider MoSe\textsubscript{2}/WSe\textsubscript{2} heterobilayers (\autoref{fig1:intro}a,b) as the prototypical TMD heterostructure for hosting IXs. The model and analysis presented here can be generalized to other heterobilayers. MoSe\textsubscript{2}/WSe\textsubscript{2} heterobilayers have a type-II staggered band alignment, in which the lowest conduction band edges and highest valence band edges are associated with MoSe\textsubscript{2} and WSe\textsubscript{2}, respectively (\autoref{fig1:intro}a). Due to the order of spin-up and spin-down states in the two types of constituent monolayers,\cite{Echeverry_Gerber_2016} the lowest IX state in a MoSe\textsubscript{2}/WSe\textsubscript{2} heterobilayer is optically active.

To reflect the associated photodynamic processes during the lifetimes of IXs, we explicitly consider exciton diffusion, recombination, trapping,  bright-dark flipping, and exciton-exciton-annihilation (EEA) on an equal footing with the formation and dipole-dipole interaction of IXs (\autoref{fig1:intro}c,d). 
To a first approximation, we neglect the influences of exciton complexes and exciton-exciton correlations.
A discrete-time random walk of fixed step size provides an excellent description of exciton diffusion phenomena while simultaneously incorporating the other statistical processes. A full explanation of the model used in this study is given in the Supporting Information along with the parameters we use to simulate a MoSe\textsubscript{2}/WSe\textsubscript{2} heterostructure (Table S1). A brief description of the model is given below and sketched in \autoref{fig1:intro}c,d.

\begin{enumerate}[wide,labelwidth=!, labelindent=10pt, label=(\roman*)]
\item
The evolution of the set of random walkers, $\mathbb{S}^t$, from time $t$ to $t+\delta t$ is defined by the updating schema 
\begin{align} \label{eq:prop}
\mathbb{S}^{t+\delta t} = \tilde{\mathfrak{P}}\left[\mathbb{S}^t\right] = \tilde{\mathfrak{p}}_i\left[\cdots \tilde{\mathfrak{p}}_b\left[\tilde{\mathfrak{p}}_a\left[\mathbb{S}^t\right]\right]\right],
\end{align}
in which $\tilde{\mathfrak{P}}$ is constructed from functional, stochastic, sub-propagators $\tilde{\mathfrak{p}}_i$ that are defined to account for physical processes such as diffusion, recombination, EEA and dipole-dipole repulsion.

\item
At each time step, movement is accomplished for each element of $\mathbb{S}$ by $\tilde{\mathfrak{p}}_{\text{move}}$
which moves each exciton a specific distance, $\delta r$, with a random angle, $\tilde{\theta}$, drawn from a uniform distribution $[0,2\pi)$
\begin{align}
\begin{pmatrix}
x \\
y
\end{pmatrix}^{t+\delta t} = \begin{pmatrix}
x \\
y
\end{pmatrix}^t +
\begin{pmatrix}
\delta r \cdot \cos{\left(\tilde{\theta}\right)} \\
\delta r \cdot \sin{\left(\tilde{\theta}\right)} 
\end{pmatrix}.
\end{align} 
The spatial stepsize is held constant and calculated from the diffusivity in two-dimensions, $D$,
\begin{align}
\delta r &= \sqrt{4D\cdot\delta t}.
\end{align}

\item
Hopping into and out of traps is controlled by diffusion and a Miller-Abrahams probability weighting factor,\cite{Miller_Abrahams_1960}
\begin{align}
	P_{\text{move}} = \exp{\left(\frac{-\Delta E}{k_BT}\right)},
\end{align} 
which yields the moving probability being dependent on the relative energies of the current and proposed new locations for each exciton. 

\item
To account for the trapping-induced changes in the recombination lifetimes, $\tau^{\text{2D}}$,\cite{Wang_Ma_2020, Peng_Ma_2020, Srivastava_Imamoglu_2015} we adapt the concept of 2D exciton coherence area which is determined by exciton scattering.\cite{Feldmann_Elliott_1987, Ma_Gosztola_2017} From this model, we find that the exciton radiative lifetime is dependent on the exciton coherence radius $r_c$, oscillator strength $f_0$, and temperature $T$, 
\begin{align}
	\tau^{\text{2D}} &= \frac{a}{ r_c^2 f_0}\left[1-\exp\left(-\frac{2\hbar^2}{M r_c^2k_BT}\right)\right]^{-1}, \label{eq:radlifetime}
\end{align}
with $a$ being a collection of constants directly related to the material, and $M$ being the exciton's reduced mass.
During our calculations, $r_c$ is set to the trap radius, $r_{\text{trap}}$, if an exciton is in a trap. The oscillator strengths are chosen to emulate radiative lifetimes found in the literature with $f_{0,\text{bright}} \ll f_{0,\text{IX}} \ll f_{0,\text{dark}}$ (see Table S1 for the detailed parameters).  

\item
The EEA process typically happens on ultrafast time scales in quantum confined semiconductors and is one of the major parasitic mechanisms in diminishing exciton populations.\cite{Mouri_Matsuda_2014, Ma_Htoon_2015} In our model, it is considered to occur when two excitons are within a certain annihilation capture radius, which results in the  stochastic annihilation of one of the excitons and the remaining one being moved to their joint Euclidean mid-point. 

\item
The strong dipole-dipole repulsive interactions $U_{\text{dd}}$ among IXs with dipole, $p$, separated by a pair distance $r$,
\begin{align}
U_{\text{dd}}(r) =- \frac{p^2}{\epsilon\epsilon_0r^3}, \label{eq:dipole}
\end{align}
are implemented by using a mean-field approach. 
The change in the dipole potential energies $\Delta U$ caused by dipole-dipole interactions as an IX moves a distance $\Delta r$
\begin{align}
\Delta U = \int_{r}^{r+\Delta r} \vec{F} \text{d}\vec{r}
\end{align}
is incorporated into the Miller-Abrahams hopping rates. 
Here, $\vec{F}$ is the mean-field force caused by all pair-potential dipole interactions, 
 
\item
In our preliminary modeling, we found that due to the ultrafast interlayer charge transfers in MoSe\textsubscript{2}/WSe\textsubscript{2} heterobilayers,\cite{Ceballos_Zhao_2014} we can treat them as unified four state systems of ground state, bright and dark intralayer exciton states, and IX state with dynamic interconnections shown in \autoref{fig1:intro}d. 

\end{enumerate}

\section{Results and Discussion}

To exemplify the workflow of our simulation, we consider an array of 5$\times$5 ($N_{\text{trap}} = 25$) flat-bottomed potential traps that are 20 nm in radius ($r_{\text{trap}} = 20$ nm) and equally separated by 50 nm (\autoref{fig1:intro}e) at 5 K. The depths of the potential traps, $E_{\text{trap}}$, are set to be 50 meV. An instantaneous, diffraction limited Gaussian pulse with a full-width-at-half-maximum of 215 nm centered at the trap array populates an initial exciton number of $N_{\text{ex},0}=1000$ into the bright intralayer exciton state. This excitation condition corresponds to an initial density of 2$\times$10\textsuperscript{12} excitons per cm\textsuperscript{2} in the diffraction-limited excitation spot. The spatial dimensions of the simulation are not bounded, however the temporal dimension is bounded to [0,5] ns (100000 steps).

\autoref{fig1:intro}f shows time slices of these photogenerated excitons. The rows correspond to different types of excitons: decayed excitons including both intralayer and interlayer excitons (``gone"), bright intralayer excitons (``bright"), dark intralayer excitons (``dark"), and interlayer excitons (``IX"). The total number of the specific exciton type still ``living" at that moment is shown in the top left corner of each frame while the number of the corresponding trapped excitons is in the top right corner. \autoref{fig1:intro}g presents the total and trapped populations of each exciton type at each time point of the simulation shown in \autoref{fig1:intro}f.

At time $t$ = 0, 1000 bright intralayer excitons are generated, followed by their recombination or diffusion (\autoref{fig1:intro}f and 1g, second row). Alternatively, the bright intralayer excitons can also convert to dark intralyer excitons (\autoref{fig1:intro}f and 1g, third row) or IXs (\autoref{fig1:intro}f and 1g, bottom row), which all eventually transition to the ground state (\autoref{fig1:intro}f and 1g, top row).
Once an exciton has converted to the ground state, it is no longer propagated. 
For simplicity, in the following, we refer to the bright (dark) intralayer excitons as the bright (dark) excitons.
While the consumption of the bright excitons occurs within the first 5 ps, consistent with previous reports,\cite{Hong_Wang_2014} the build up of free and trapped dark and interlayer excitons and their subsequent decay happens over hundreds of picoseconds (\autoref{fig1:intro}g). 
Diffusion of the free IXs is clearly evident in the evolution of the IX frames of \autoref{fig1:intro}f (bottom row); these frames track the locations of the IXs before they decay to the ground state.

In order to quantify the trapping efficiency for a given set of simulation conditions, we define a metric, 
\begin{align}
\text{trapping efficiency}_{\text{IX}} \equiv \eta_{\text{IX}} \equiv \frac{\int{N_{\text{IX, trap}}(t)\text{d}t}}{\int{N_{\text{IX, total}}(t)\text{d}t}}, \label{eq:trapratio}
\end{align}
in which $N_{\text{IX, total}}(t)$ and $N_{\text{IX, trap}}(t)$ are the total and trapped number of ``living" IXs, respectively, at a given simulation time point $t$. Their time evolution curves are shown in  \autoref{fig1:intro}g (bottom row).
$\eta_{\text{IX}}$ is unity when all IXs are trapped during the entire simulation and less than unity when some IXs are not trapped.
This value gives a time- and trap-averaged evaluation of the trapping capability throughout the lifetimes of the IXs.
Note that this metric can be applied to other types of excitons, but in the main text we exclusively apply it to the IXs. 
Another important measure of the trapping behavior is the average number of IXs localized per potential trap before they undergo recombination, $n_\text{IX,trap}$. Since an IX may experience multiple trapping and detrapping throughout its lifetime, we use the maximum number of trapped IXs divided by the trap number as an estimation of $n_\text{IX,trap}$. For the specific example shown in \autoref{fig1:intro}g (bottom row) we find $n_\text{IX,trap} \approx 4$. 
In the following, we systematically investigate the influence of different parameters, including the trap geometries and temperature, on the trapping properties of the IXs.

\subsection{Influence of the potential trap size}
\begin{figure*}[!htbp]
	\centering
	\iffigures
	\includegraphics[trim=.5cm 1cm 1.5cm 1.5cm, clip,width=\linewidth]{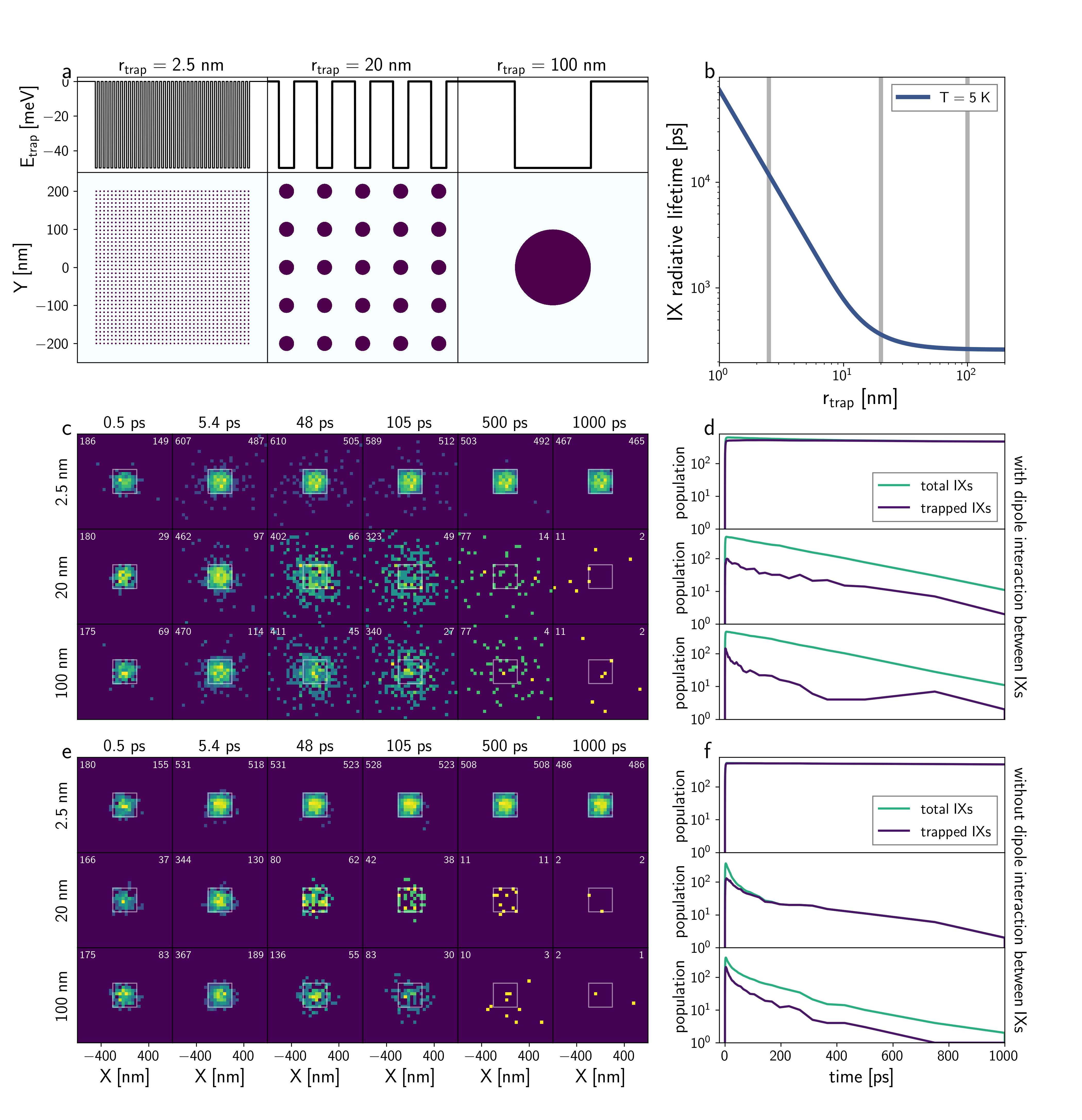}
	\fi
	\caption{ 
		Influence of trap radius on dynamics and diffusion of IXs.
		(a) Profiles of the three potential arrays with 2.5, 20, and 100 nm radii traps. All arrays have same total area, $\pi(100 \text{ nm})^2$. Top (Bottom) row: side/energetic (top) views of the potential arrays.
		(b) IX lifetime vs. trap radius at 5 K from \autoref{eq:radlifetime} on a log-log scale; vertical gray bars correspond to the trap radii of (a).
		(c, e) Evolution of IXs through space and time for different sized potential traps with (c) and without (e) considering the dipole-dipole repulsion term. $T=5\text{ K}$, $E_{\text{trap}}=50\text{ meV}$, and $N_{\text{ex},0}=1000$.   
		The convex hull of the traps is boxed in white.
		(d, f) Time evolution of the trapped and total IX populations constructed from the time frames in (c) and (e) on a log-linear scale.
	}\label{fig2:trapshape} 
\end{figure*}

Since the size of the potential traps directly defines the inter-exciton distances and their interactions, which can be manifested as the number of excitons per trap, we start by investigating the influence of the trap size on the IX trapping properties. 
During the simulation, we constrain ourselves to conditions which are experimentally relevant, \textit{i.e.} trap sizes that are comparable to those created by strains, defects, and moir\'{e} potentials.\cite{Darlington_Schuck_2020, Rosenberger_Jonker_2019, Wu_MacDonald_2018, Shanks_Schaibley_2021}
Five potential landscapes of trap arrays with radii of 100 nm, 20 nm, 11 nm, 5 nm and 2.5 nm are constructed (\autoref{fig2:trapshape}a). 
For ease of comparison, all of these primary trap potentials have the same total area. Hence, a total of $N_\text{trap}$ = 1, 25, 81, 400, and 1600 traps are simulated, respectively.
We conceptualize the different trap radii as moving from microscopic traps which can be created by strains\cite{Wang_Ma_2020, Kremser_Finley_2020} to nanoscopic traps caused by atomic defects or moir\'e potentials.\cite{Yuan_Huang_2020, Li_Srivastava_2020} Aside from the trap size, all the other simulation conditions are kept constant: temperature $T$ = 5 K, trap depth $E_{\text{trap}}$ = 50 meV and initial exciton population $N_{\text{ex},0}=1000$. Since under these conditions, the trap shape has no pronounced influence on the IX trapping behavior (Figure S3), we focus our discussion on the flat-bottom shaped potential traps.

\begin{figure}[!htbp]
	\centering
	\iffigures
	\includegraphics[trim=0.cm 2.5cm 2cm 1.5cm, clip,width=\linewidth]{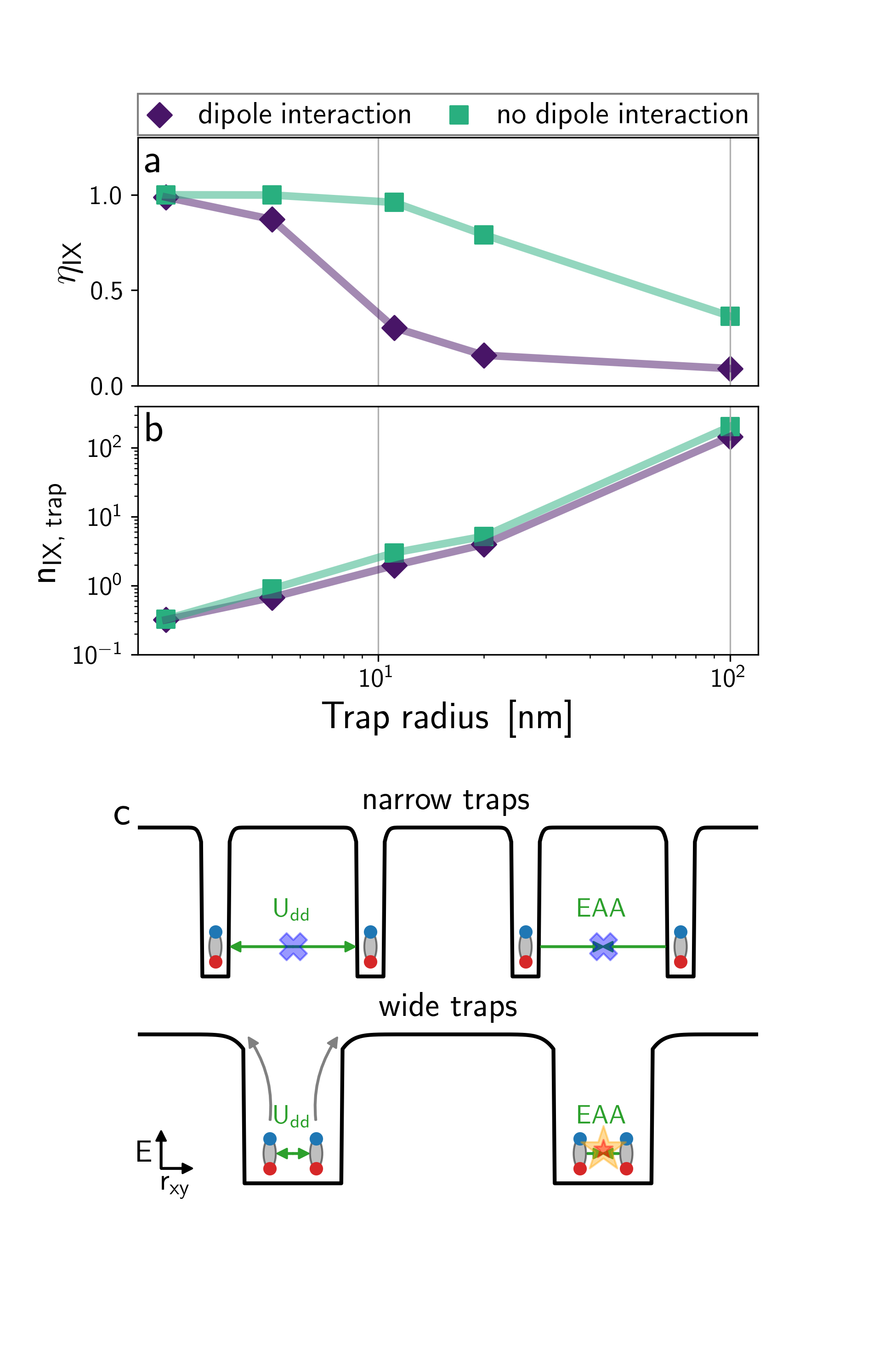}
		\fi
	\caption{
		Influence of trap radius on trapping of IXs.
		(a, b) IX trapping efficiency (a, linear-log scale)  and average number of IXs localized per potential trap, $n_\text{IX,trap}$, (b, log-log scale) vs. trap radius.
		Results are shown when dipole-dipole interactions are both on and off.
		The total area of traps, $\pi(100 \text{ nm})^2$, is constant for all simulations while $T=5\text{ K}$, $E_{\text{trap}}=50\text{ meV}$, and $N_{\text{ex},0}=1000$.
		(c) Sketch showing that narrow traps accommodating isolated IXs can effectively eliminate dipolar repulsion and exciton-exciton annihilation (top row). In contrast, wide traps that can host multiple IXs can result in dipolar repulsion pushing IXs out of the traps or exciton-exciton annihilation reducint trap population (bottom row).   
	}\label{fig3:radius} 
\end{figure}

\autoref{fig2:trapshape}c and \autoref{fig2:trapshape}d shows the evolution of the ``living" IXs in three potential trap arrays with various radii before they decay to the ground state. An apparent reduction in the lifetimes of the trapped IXs can be observed when the trap size increases from 2.5 nm to 100 nm. We believe the reason for the observed lifetime shortening is two fold. Firstly, as the size of the potential trap increases, the extent of quantum confinement experienced by the trapped IXs in the lateral dimension reduces. From a trap size of 2.5 nm to 100 nm, the quantum confinement transitions from two-dimensional to one-dimensional, considering that the Bohr radius of the IXs is a few nanometers.\cite{Li_Srivastava_2020}. By adopting an exciton coherent motion area model that have been widely used to describe two-dimensional (2D) excitons,\cite{Feldmann_Elliott_1987, Ma_Gosztola_2017} in which the transition oscillator strength of an exciton is assumed to be equivalent to the sum of the oscillator strengths of all the unit cells that contribution to the optical transitions, we find that the recombination lifetime of the trapped IXs reduces with increasing trap size (\autoref{fig2:trapshape}b). 
Secondly, an increase in the trap size may result in multiple IXs in the same potential trap, hence increasing the possibility of the EEA process. This scenario is supported by evaluating the average number of IXs localized per potential trap, $n_\text{IX,trap}$. 
As shown in \autoref{fig3:radius}b, the value of $n_\text{IX,trap}$ increases from an average of less than one IX per trap for the 2.5 nm sized traps to more than 100 IXs for the 100 nm sized traps. Apparently, IXs trapped in the latter would experience EEA processes more frequently.

To understand the trap size-dependent IX localization behavior, we further inspect the trapping efficiency $\eta_{\text{IX}}$ of the IXs as a function of the trap size (\autoref{fig3:radius}a). While an array of 2.5 nm sized traps has a near unity trapping efficiency, that of the 100 nm sized trap is decreased by an order of magnitude to only around 10$\%$. 
The most significant reduction occurs when the trap size is between 5 nm to 20 nm.
This is in stark contrast to the trapping efficiency of the dark excitons, which exhibit respectable trapping efficiencies for all of the various sized traps (Figure S5). Because of the long lifetimes of the IXs and dark excitons, it is temping to think of the IXs merely as long-lived intralayer excitons similar to dark excitons. However, the characteristic out-of-plane permanent dipole moment and the resultant dipole-dipole repulsion of the IXs distinguish them from intralayer excitons. We believe these unique features of the IXs are the primary reasons for the drastically different trapping efficiencies of the IXs and dark excitons despite their similar long lifetimes. 

\autoref{fig2:trapshape}e,f presents the evolution of the IXs when the dipole-dipole repulsion term is turned off in the simulation.
The behavior of the 2.5 nm trap simulation with dipole interactions on and off are nearly identical. Conversely, the larger trap systems have far fewer IXs at times greater than a few ps when the dipole interaction is removed. Without the dipole interaction, the IXs that do exist undergo far less stochastic movement (diffusion).
\autoref{fig3:radius}a shows that removing the dipole interaction leads to much greater trapping efficiency for large traps, however, the maximum number of excitons per trap, \autoref{fig3:radius}b, is only slightly increased by turning off the dipole interaction.
In other words, the dipole interaction enables IXs in systems with large traps to remain trapped for longer times but does not change the maximum number of IXs in traps around a picosecond after excitation. 

Based on the fact that reducing the trap size drastically increases trapping efficiencies of the IXs by taking into account dipole-dipole repulsion (\autoref{fig3:radius}a), we speculate that with myriad narrow traps, each IX has its own trap and is therefore isolated from inter-IX interactions like dipole repulsion and EEA.
However, when wide traps are present, multiple IXs congregate in each trap and they can either push each other out of the trap via dipolar interactions or the total lifetime of the ensemble can be decreased because of EEA process. 
We confirm this speculation by simulating the time evolution of the number of IXs in each trap (Figure S10). While a significant portion of the 20 nm sized traps can host more than 10 IXs during the initial few picoseconds after the photoexcitation, the number of IXs trapped in the 2.5 nm sized traps is predominantly one or none. Additionally, Figure S10 and S11 demonstrate that when a 2.5 nm trap becomes doubly occupied, the next time step it is not occupied at all--indicating that doubly occupied traps enhance EEA and IX ejection via the dipole interaction.   
We schematically illustrate these ideas in \autoref{fig3:radius}c. 

These findings explain the sudden start of reduction in the IX trapping efficiency when the dipole-dipole repulsion is included and the trap size is between 5 nm to 20 nm (\autoref{fig3:radius}a): the dipole-dipole repulsion is most effective when the inter-IX distance is a few nanometers due to the $1/r^4$ scaling of the dipole force (Equation S34).
More importantly, these results underscore the optimal conditions for localizing individual IXs in potential traps: it is critical to have small traps that are a few nanometers in size. This finding agrees well with the recent observations of single IXs localized in a moir\'e potential trap\cite{Li_Srivastava_2020} and multiple IXs in strain-induced traps\cite{Wang_Ma_2020}: while both types of potential traps may offer similar trapping depths, the moir\'e potential traps are typically a few nanometers in size,\cite{Li_Srivastava_2020,Yuan_Huang_2020} significantly smaller than the strain-induced traps\cite{Rosenberger_Jonker_2019}, and thus allow the localization of individual IXs.

\subsection{Influence of the exciton density}

We further interrogate the influence of exciton density on the trapping behavior of the IXs. For ease of comparison, we keep the system temperature at 5 K, the trap depth at 50 meV, and the trap radius at 20 nm. The initial bright, intralayer exciton population $N_{\text{ex},0}$ is varied between 125 and 4000, while the number of traps $N_{\text{trap}}$ is tuned from 4 to 36. 
In this way, we constrain the initial exciton density to below 10\textsuperscript{13} cm\textsuperscript{-2} because at higher densities, the free IX picture may no longer be valid and instead, degenerate electron-hole plasmas can form.\cite{Wang_Zhu_2021}  

\begin{figure}[!htbp]
	\centering
	\iffigures
	\includegraphics[trim=0.5cm 1.cm 2.cm 2cm, clip,width=\linewidth]{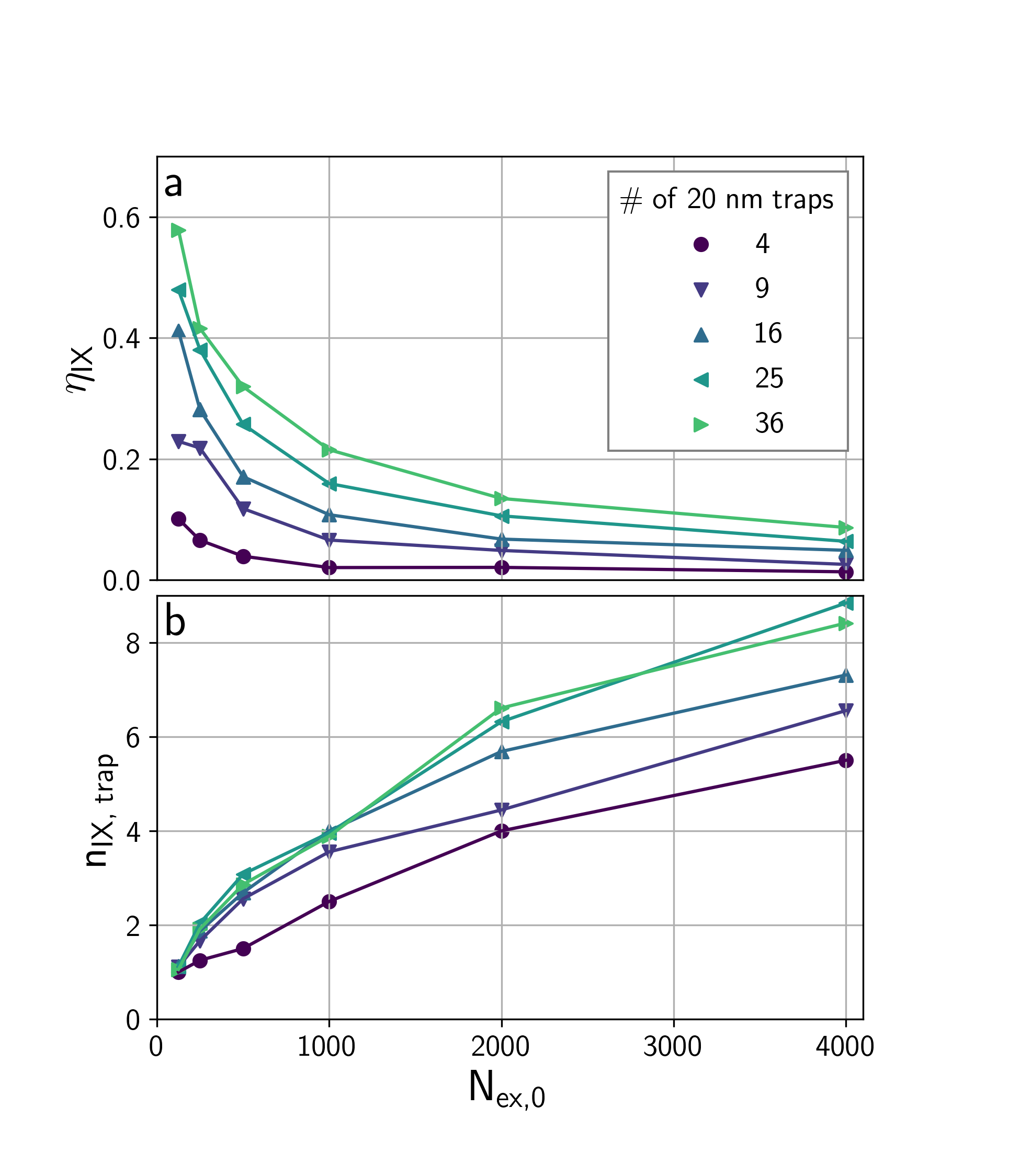}
	\fi
	\caption{
		Influence of trap and exciton density on trapping of IXs in 20 nm traps.
		(a, b) IX trapping efficiency (a)  and average number of IXs localized per potential trap, $n_\text{IX,trap}$, (b) vs. initial bright exciton densitys.
		In all simulations, $r_{\text{trap}} = 20 \text{ nm}$, $T=5\text{ K}$, and $E_{\text{trap}}=50\text{ meV}$.  
		Different numbers of traps are represented as different colors and markers.
	}\label{fig4:density} 
\end{figure}

\autoref{fig4:density}a shows that a reduction in the initial exciton density (i.e. exciton population per trap), whether through decreasing the initial exciton population $N_{\text{ex},0}$ or increasing the trap number $N_{\text{trap}}$, leads to an increase in the IX trapping efficiency. Similar results can be observed for 2.5 nm sized potential traps (Figure S8), although the relative trapping efficiency is considerably higher. 
We believe the reason for this exciton density-dependent trapping behavior is the same as that shown in \autoref{fig3:radius}c: populating the same potential trap with multiple IXs would result in their strong repulsive interactions. As a consequence, some of the IXs get pushed out of the traps, leading to less efficient trapping at higher exciton densities.
Additionally, \autoref{fig4:density}b shows that increasing the initial exciton density causes $n_\text{IX,trap}$ to increase. The relationship between $n_\text{IX,trap}$ and initial density is sub-linear at large densities indicating a significant saturation behavior. Seemingly, a larger number of traps moves the saturation onset to larger densities.
These observations are consistent with previously observed exciton density-dependent IX transport, in which the effect of the moir\'{e} potential was screened at high exciton densities.\cite{Yuan_Huang_2020, Seyler_Xu_2019}
Indeed, \textcite{Wang_Zhu_2021} experimentally observed a moir\'e trapped-IX to free-IX gas transition in  MoSe\textsubscript{2}/WSe\textsubscript{2} at IX densities (10\textsuperscript{11} cm\textsuperscript{-2}) in the same range which we investigate here.  


\subsection{Influence of temperature and trap depth}

One of the most appealing properties of the IXs in TMDs when compared to indirect excitons in traditional semiconductor heterostructures is their large binding energy,\cite{Gillen_Maultzsch_2018,Donck_Peeters_2018} which may open many opportunities for TMD-based quantum photonic and optoelectronic devices operating at elevated temperatures.  
For example, quantum photon sources that function at room temperature may offer many technical advantages.\cite{Tran_Aharonovich_2016,Ma_Htoon_2015_B} Recent experimental and theoretical studies have suggested the possibility of realizing high temperature exciton condensation and superconductivity based on IXs in TMDs.\cite{Wang_Mak_2019,Fogler_Novoselov_2014, Lu_Ye_2018}
An understanding of temperature- and trap depth-dependent IX trapping in TMD heterostructures is of high practical relevance.

\begin{figure*}[!htbp]
	\centering
	\iffigures
	\includegraphics[trim=.5cm 1cm 2.0cm .5cm, clip,width=\linewidth]{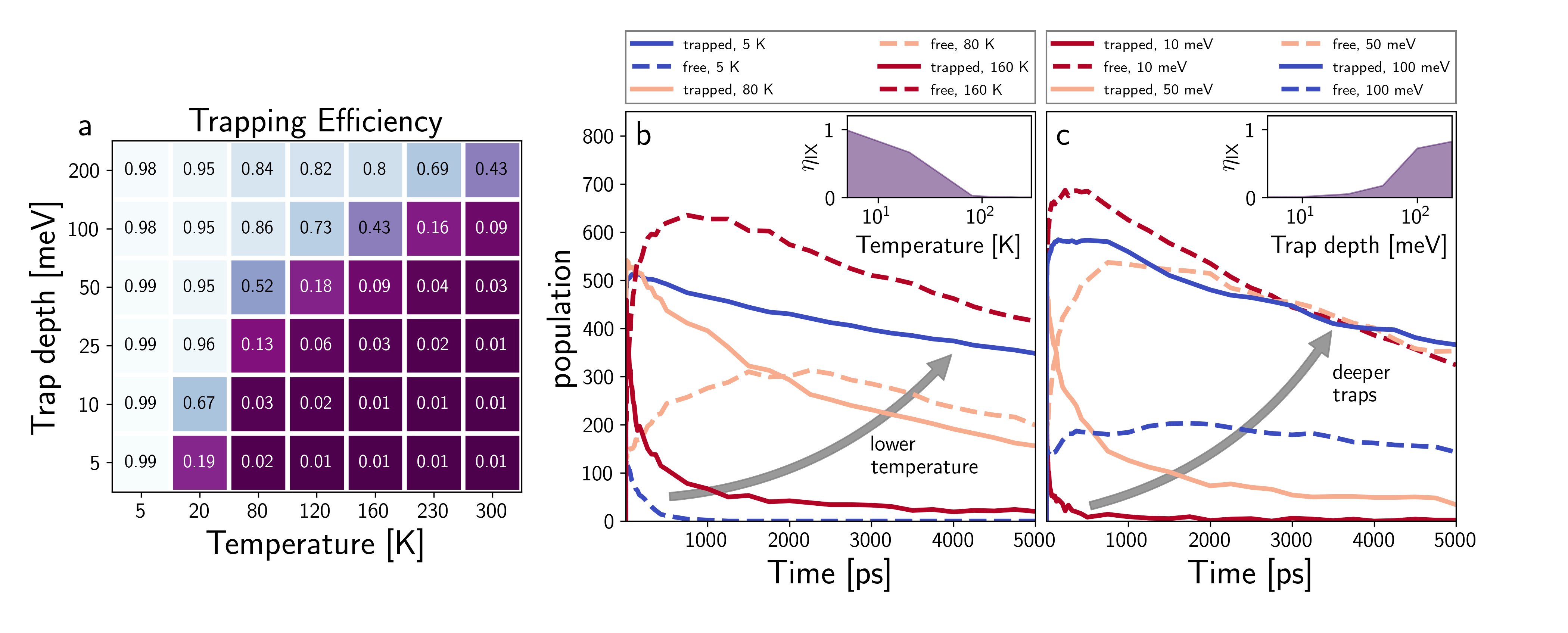}
	\fi
	\caption{ 
		Influence of temperature and trap depth on dynamics and trapping efficiency of IXs in 1600 2.5 nm traps. 
		(a) Trapping efficiencies when trap depth spans 5 to 200 meV and from 5 to 300 K with a lighter color maps to higher efficiency. 
		(b, c) Free (dashed) and trapped (solid) IX population vs. time.
		(b) 5, 80, and 160 K simulations with 50 meV deep traps; the inset graphs trapping efficiency vs. temperature for 50 meV deep traps. 
		(c) 5, 50, and 100 meV traps at 120 K; the inset graphs trapping efficiency vs. trap depth at 120 K. 
		As indicated by the gray arrows, deeper traps and lower temperatures lead to larger trapping efficiency and longer living trapped IXs. 
	}\label{fig5:energetics} 
\end{figure*}

The influences of the system temperature and trap depth are investigated by varying the former in the range of 5 - 300 K and the latter in the range of 5 - 200 meV while maintaining the same trap area and bright exciton density of 1000. This trap depth range is representative of experimentally obtainable values: strain-induced traps typically have depths of a few tens to a few hundreds of meV,\cite{Darlington_Schuck_2020, Rosenberger_Jonker_2019, Peng_Ma_2020, Branny_Gerardot_2017} while that of moir\'{e} potentials can be a few hundreds of meV.\cite{Zhong_Shih_2017}
We find that increasing the system temperature reduces the trapping efficiency (\autoref{fig5:energetics}a). A similar reduction phenomenon can be observed when the trap depth is reduced, although to a much lesser extent. 
For an increase in the trap depth, a consistent increase in the trapping efficiency can be observed until a plateau is reached, which defines the maximum achievable IX trapping efficiency by optimizing the trap depth while keeping the other parameters the same.
A similar trend is observed for 20 nm radii traps (Figure S6b).

For 50 meV deep 2.5 nm radii traps, \autoref{fig5:energetics}b shows that at 5 K trapped IXs have a far greater population than free IXs. As the temperature is increased, the dominant population flips between 80 and 160 K. This change in dominant species is analogous to the trapped-to-free-gas transition observed by \textcite{Wang_Zhu_2021}.
Fixing the temperature at 120 K and changing the trap depth (\autoref{fig5:energetics}c) demonstrates that a similar flip between the dominant population happens at a trap depth between 50 and 100 meV. 
The thermodynamic controls of temperature and trap depth therefore not only have jurisdiction over temporally integrated behavior like trapping efficiency (\autoref{fig5:energetics}a, which relate to steady-state device performances); these controls also have a rich interplay with each other to define the ultrafast evolution of the IX populations and whether the dominant species is trapped or a freely diffusing gas.

\section{Conclusions}
Using a discrete-time random walk model, we study the trapping behavior of IXs in TMD heterostructures. By systematically investigating the influences of trap parameters such as size and depth and externally controlled parameters like system temperature and exciton density, we are able to illustrate conditions when IX localization should be expected. We find that dipolar repulsion among IXs distinguishes them from intralayer excitons and has a strong impact on their trapping. This is manifested as low trapping efficiencies of IXs in large traps, in which dipole-dipole repulsion helps push IXs out of the traps. In contrast, nanoscale traps afforded by pointed defects and moir\'e potentials are most effective in localizing the IXs---high densities of trapped IXs are achievable when sufficient deep traps are provided for each IX to be separated from its neighbor.
The same mechanism is also reflected in the low trapping efficiencies at high exciton densities, where increases in dipolar repulsive interactions effectively screen the effect of the potential traps. 
We also show that temperature and trap depth form a rich control system for defining whether the dominant IX population is trapped or free.
These results indicate the importance of engineering nanoscale traps such as point defects and moir\'e potentials in order to effectively trap IXs for quantum photonic applications.

\section{Supporting Information}

The Supporting Information is available free of charge at \textbf{URL} and includes 
computational methods,
results demonstrating the role of trap shape and dark excitons,
and a discussion of individual trap occupancy.

\begin{acknowledgements}
This work was performed, in part, at the Center for Nanoscale Materials, a U.S. Department of Energy Office of Science User Facility, and supported by the U.S. Department of Energy, Office of Science, under Contract No. DE-AC02-06CH11357. X.M. acknowledges support by the Department of Energy, Center for Molecular Quantum Transduction Energy Frontier Research Center, under Grant No. DE-SC0021314. 
\end{acknowledgements}

\bibliography{database}

\begin{figure*}[htbp]
	\centering
	\iffigures
	\includegraphics[width=4.45cm]{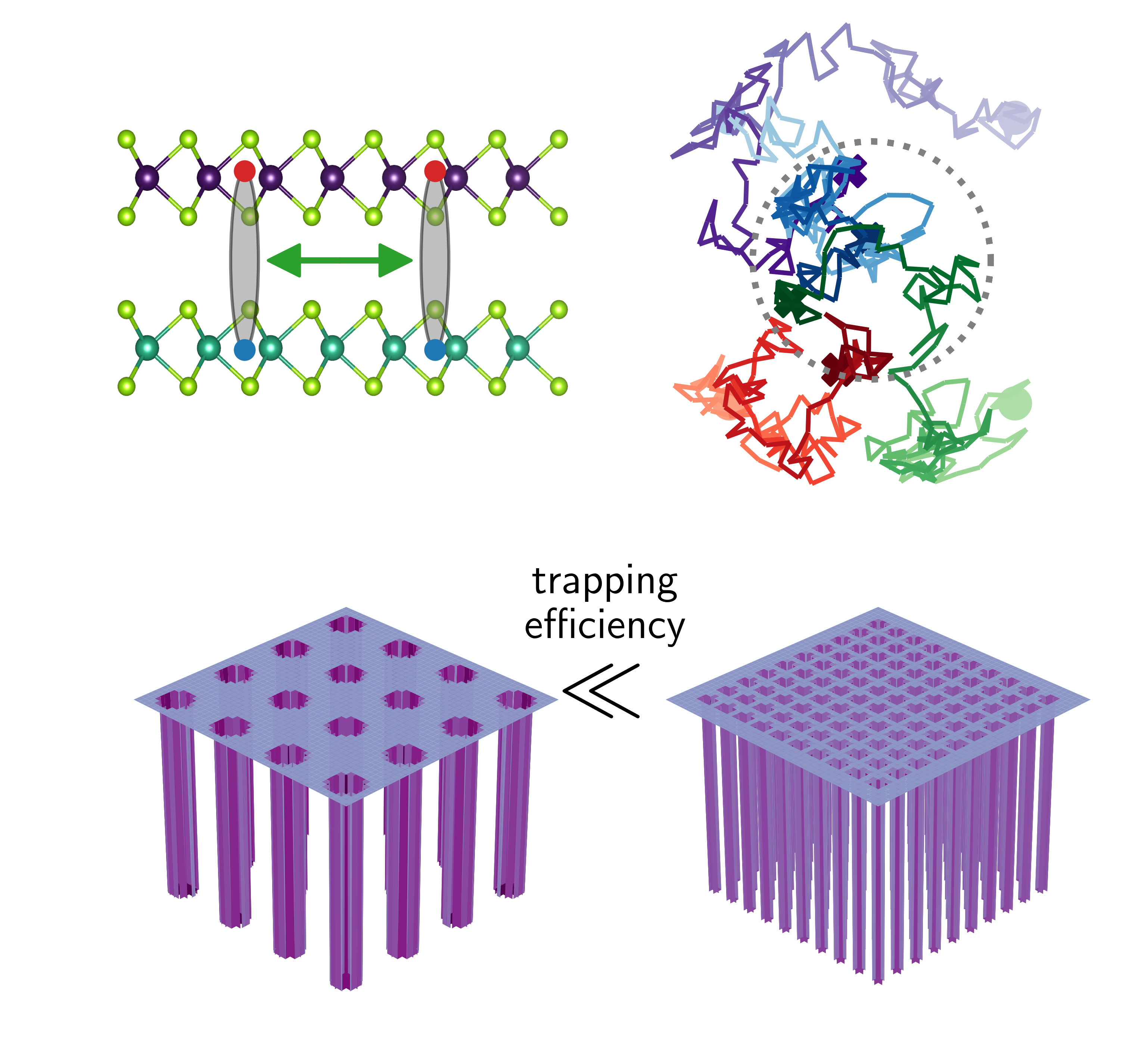}
	\fi
	\caption{
		For Table of Contents Only
	}\label{figTOC} 
\end{figure*}

\end{document}


\title{Supplementary Material: Trapping Interlayer Excitons in van der Waals Heterostructures by Potential Arrays}

\author{Darien J. Morrow}
\affiliation{Center for Nanoscale Materials, Argonne National Laboratory, Lemont, Illinois 60439, United States}

\author{Xuedan Ma}
\email{xuedan.ma@anl.gov}
\affiliation{Center for Nanoscale Materials, Argonne National Laboratory, Lemont, Illinois 60439, United States}
\affiliation
{Consortium for Advanced Science and Engineering, University of Chicago, Chicago, Illinois 60637, United States}
\affiliation
{Northwestern-Argonne Institute of Science and Engineering, 2205 Tech Drive, Evanston, IL 60208, USA}

\date{\today}

\maketitle
\tableofcontents
\clearpage
\section{Computational methods}

In this work we desire to explicitly treat conversion among exciton types, trapping, annihilation, and dipolar repulsion of excitons on an equal footing with their diffusive behavior.
These desired behaviors are not trivial to add to the normal, Fickian continuum model of diffusion with drift coupled to linear (radiative recombination) and quadratic (exciton-exciton annihilation) loss terms.
Hence, we developed a \emph{discrete} model of exciton dynamics on a plane; this model is a progeny of both Random Walk and Markov Chain models.

\subsection{Bookkeeping: the structure of the simulation}
The entire simulation universe, $\mathbb{S}^t$, at a given time, $t$, is defined as the tuple of all excitons
\begin{align}
\mathbb{S}^t \equiv \left(\text{ex}_1^t, \text{ex}_2^t, \dots, \text{ex}_n^t\right)
\end{align}
The $n$th exciton is entirely defined at $t$ by the tuple
\begin{align}
\text{ex}_n^t \equiv \left(x, y, \mathbb{I}\right),
\end{align} 
in which $x$ and $y$ are Cartesian spatial coordinates and $\mathbb{I}$ is the \emph{identity} of the exciton at that time.
The values that $\mathbb{I}$ can take depend on the specific simulation. 
For most of the simulations in this work
\begin{align}
\mathbb{I} \in \left\{\text{gone},\text{bright},\text{dark},\text{interlayer}\right\}.
\end{align}

The simulation is advanced a timestep, $\delta t$, by a propagator, $\tilde{\mathfrak{P}}$, such that
\begin{align}
\mathbb{S}^{t+\delta t} = \tilde{\mathfrak{P}}\left[\mathbb{S}^t\right], \label{eq:prop}
\end{align}
in which the tilde notation means some aspect of the propagator uses a random number generator (this will be more concretely detailed below). 
Much like a Markov Chain model, this simulation is memory-less because $\mathbb{S}^t$ is uniquely determined (within random number sampling) by $\mathbb{S}^{t-\delta t}$. 

$\tilde{\mathfrak{P}}$ is constructed to provide all physics needed for the problem of interest.
Specifically, functional sub-propagators, $\tilde{\mathfrak{p}}_i$, are defined to account for physics like movement, decay, annihilation, repulsion, etc. 
For $i$ sub-propagators, \autoref{eq:prop} is rewritten as nested functionals
\begin{align}
\mathbb{S}^{t+\delta t} &= \tilde{\mathfrak{P}}\left[\mathbb{S}^t\right] \\
&= \tilde{\mathfrak{p}}_i\left[\cdots \tilde{\mathfrak{p}}_b\left[\tilde{\mathfrak{p}}_a\left[\mathbb{S}^t\right]\right]\right].
\end{align}
Note that the ordering of the $\tilde{\mathfrak{p}}_i$ when composing  $\tilde{\mathfrak{P}}$ is arbitrary and could subtly influence the outcome of a simulation because propagators are not necessarily linear operators.

Computationally, $\mathbb{S}$ is constructed as a NumPy array,\cite{Harris_Oliphant_2020} with sub-propagators being written in either native-NumPy or in Python and then just-in-time-compiled using Numba.\cite{Lam_Seibert_2015}

\subsection{Movement and traps}

At each time step, movement is accomplished for each element of $\mathbb{S}$ by $\tilde{\mathfrak{p}}_{\text{move}}$
which moves each exciton a specific distance, $\delta r$, with a random angle, $\tilde{\theta}$, drawn from a uniform distribution $[0,2\pi)$
\begin{align}
\begin{pmatrix}
x \\
y
\end{pmatrix}^{t+\delta t} = \begin{pmatrix}
x \\
y
\end{pmatrix}^t +
\begin{pmatrix}
\delta r \cdot \cos{\left(\tilde{\theta}\right)} \\
\delta r \cdot \sin{\left(\tilde{\theta}\right)} 
\end{pmatrix}.
\end{align} 
For a discrete 2D system, the diffusivity, $D$, is given by
\begin{align}
D&=\frac{(\delta r)^2}{4\delta t} \\
\Longrightarrow \delta r &= \sqrt{4D\cdot\delta t}
\end{align}
which we take to be the definition of the spatial stepsize $\delta r$ for each timestep. 

\begin{figure}[!htbp]
	\centering
	\includegraphics[trim=0 1cm 0 1cm, clip, width=\linewidth]{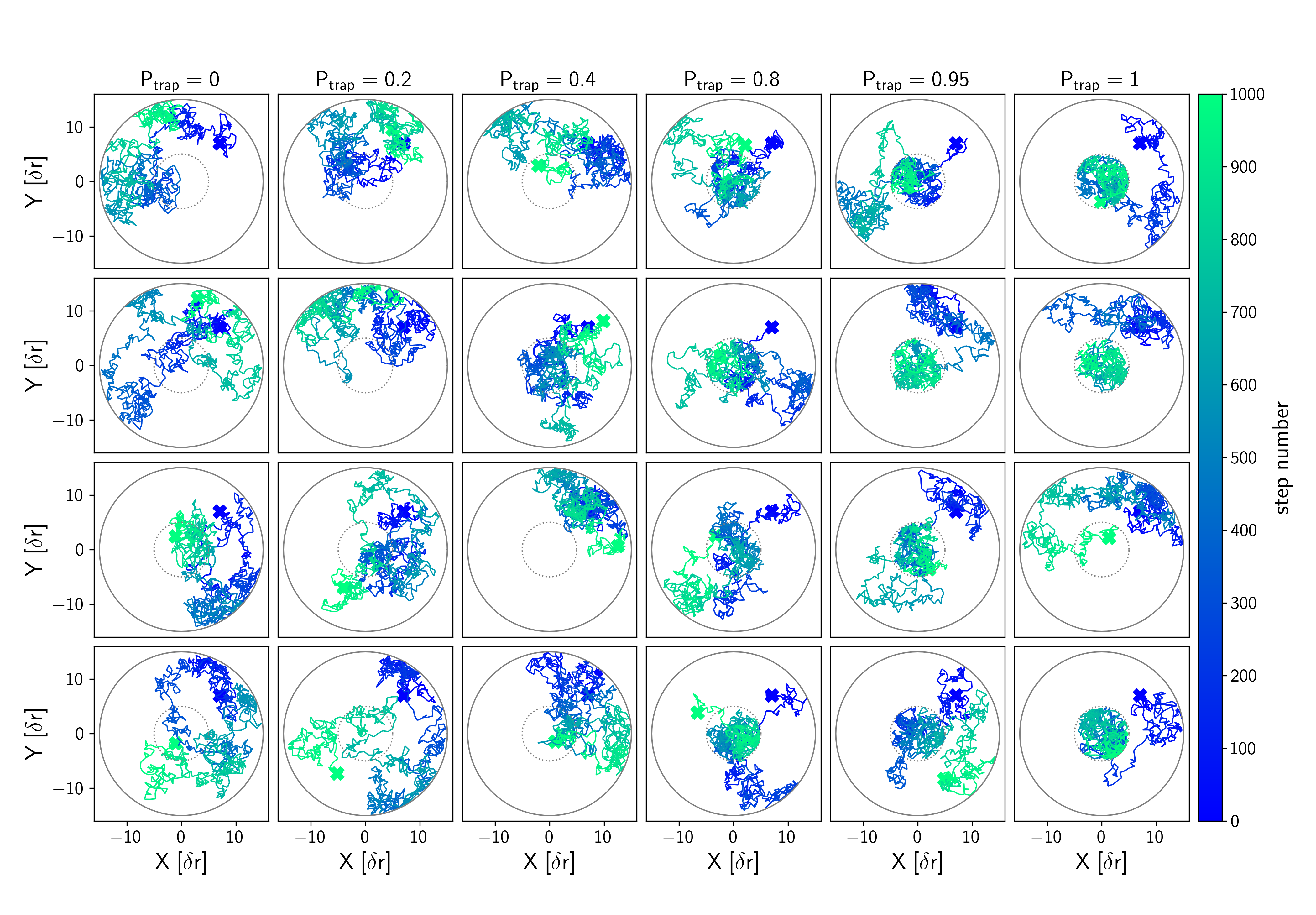}
	\caption{
		Example random walking and trapping for single walkers. Each column has a different probability of staying inside the trap once inside. The radius of the trap is $5\cdot\delta r$, there is also a hard boundary on the simulation with radius $15\cdot\delta r$. The four rows are each an iteration of the same simulation showing how the evolution of the walkers is dependent on the random number generator. The line color is mapped to simulation step number. Thick Xs mark the beginning and end points after 1000 steps.
	}\label{fig:random_walk} 
\end{figure}

Spatial traps are implemented with a Miller-Abrahams hopping probability with the potential energy surface of the system given by $E(x,y)$.  
After a new position is calculated for each particle,$\left(x_{\text{calc}}^{t+\delta t}, y_{\text{calc}}^{t+\delta t}\right)$, a probability $P_{\text{move}}$ for moving to that new position is calculated
\begin{align}
P_{\text{move}} =
\begin{cases}
\exp{\left[\frac{E\left(x^{t}, y^{t}\right)-E\left(x_{\text{calc}}^{t+\delta t}, y_{\text{calc}}^{t+\delta t}\right)}{k_B T}\right]},& \text{if } E\left(x_{\text{calc}}^{t+\delta t}, y_{\text{calc}}^{t+\delta t}\right) > E\left(x^{t}, y^{t}\right)\\
1,& \text{if } E\left(x_{\text{calc}}^{t+\delta t}, y_{\text{calc}}^{t+\delta t}\right) \leq E\left(x^{t}, y^{t}\right).
\end{cases} \label{eq:hoprate}
\end{align}
Then for a random number, $\tilde{P}_{\text{choice}}$ drawn from a uniform distribution $[0,1)$
\begin{align}
\begin{pmatrix}
x \\
y
\end{pmatrix}^{t+\delta t} =
\begin{cases}
\begin{pmatrix}
x \\
y
\end{pmatrix}^{t+\delta t}_{\text{calc}}, & \text{if } P_{\text{move}} \geq \tilde{P}_{\text{choice}}\\
\begin{pmatrix}
x \\
y
\end{pmatrix}^{t}, & \text{if } P_{\text{move}} < \tilde{P}_{\text{choice}}.
\end{cases} \label{eq:hopprob}
\end{align}
Taken together, \autoref{eq:hoprate} and \autoref{eq:hopprob} say that the probability of moving from one point to another is unity if the new point is lower in energy than the current point, but if the new point is higher in energy, then the probability of moving is given by a Boltzmann factor.
\autoref{fig:random_walk} shows an example of movement and trapping for different trap depths.

\subsection{Identity exchange}

Recombination, Bright-Dark flipping, interlayer exciton formation, etc are all accounted for by using an identity exchange step, $\mathfrak{p}_{\text{exch}}$.
Unlike a Markov chain which relies on occupation probabilities, at the end of a step, all excitons must be in a definite identity state. 
An exchange matrix, $\mathbf{P}_{\text{exch}}$, is constructed which specifies for all possible initial states $m_i$ the probability of changing to another state $n_f$ per simulation step. 
An example of such a matrix for a system which has three possible states is
\begin{align}
\mathbf{P}_{\text{exch}} = 
\begin{blockarray}{c@{\hspace{1pt}}rrr@{\hspace{3pt}}}
& a_i   & b_i   & c_i \\ \cline{2-4}
\begin{block}{r@{\hspace{1pt}}
	@{\hspace{1pt}}|lll@{\hspace{1pt}}}
a_f & P_{a\rightarrow a} & P_{b\rightarrow a} & P_{c\rightarrow a} \\
b_f & P_{a\rightarrow b} & P_{b\rightarrow b} & P_{c\rightarrow b} \\
c_f & P_{a\rightarrow c}   & P_{b\rightarrow c}   & P_{c\rightarrow c}   \\
\end{block}
\end{blockarray}.
\end{align}
Each column of $\mathbf{P}_{\text{exch}}$ must sum to unity,
\begin{align}
\sum_{f} P_{i\rightarrow f} = 1, \label{eq:unityprob} 
\end{align}
to ensure that the simulation is not an open system. 
Operationally, all probabilities are specified other than the diagonal terms which are then calculated on the fly to ensure that \autoref{eq:unityprob} is satisfied
\begin{align}
P_{i\rightarrow i} = 1 - \sum_{f\neq i} P_{i\rightarrow f}.
\end{align}

The way that $\mathfrak{p}_{\text{exch}}$ advances the simulation is to operate on all identity values of $\mathbb{S}$ 
\begin{align}
\mathbb{I}^{t+\delta t} = \mathfrak{p}_{\text{exch}}\left[\mathbb{I}^{t}\right].
\end{align}
Notationally we say $\mathbb{I}\in \left\{a,b,c,\dots, m,n\right\}$ and $\mathbb{I}^t = i$. 
For a random number, $\tilde{P}_{\text{choice}}$ drawn from a uniform distribution $[0,1)$, the value of $\mathbb{I}^{t+\delta t}$ is given by 
\begin{align}
\mathbb{I}^{t+\delta t} = m \text{ such that }\sum_{f=0}^{m-1} P_{i\rightarrow f} < \tilde{P}_{\text{choice}} \leq \sum_{f=0}^m P_{i\rightarrow f}. 
\end{align}

\subsubsection{Connection of transfer probabilities to observables}

In order to connect to experimental observables, the $ P_{i\rightarrow f}$ must be connected to rate constants.
For a single process, the probability per time step to leave a state for another state is $P_\ell$ with the probability to stay in the state being $P_s=1-P_\ell$.
The probability to be in a state after $\alpha$ steps of temporal length $\delta t$ is 
\begin{align}
P(t=\alpha\cdot\delta t) = \left(P_s\right)^\alpha
\end{align}
with the connection to a continuous decay rate being derived from the well known case of compound interest
\begin{align}
k = \frac{1}{\tau} = \frac{\ln{\left(P_s\right)}}{\delta t}.
\end{align}
So the probability per simulation step to move from an intial state to a final state with a specific rate constant, $k_{i\rightarrow f}$, is
\begin{align}
P_{i\rightarrow f} = 1 - \exp\left(-k_{i\rightarrow f} \delta t\right)
\end{align} 
Many of the coupled states in the simulations are assumed to be in thermal equilibrium, so the reverse probability is given by a Boltzmann factor
\begin{align}
P_{b\rightarrow a} = P_{a\rightarrow b} \cdot \exp\left(\frac{E_a - E_b}{k_B T}\right)
\end{align} 

\subsubsection{Radiative rates}

\begin{figure}[!htbp]
	\centering
	\includegraphics[trim=0.cm 1.cm 0.cm 2cm, clip,width=0.7 \linewidth]{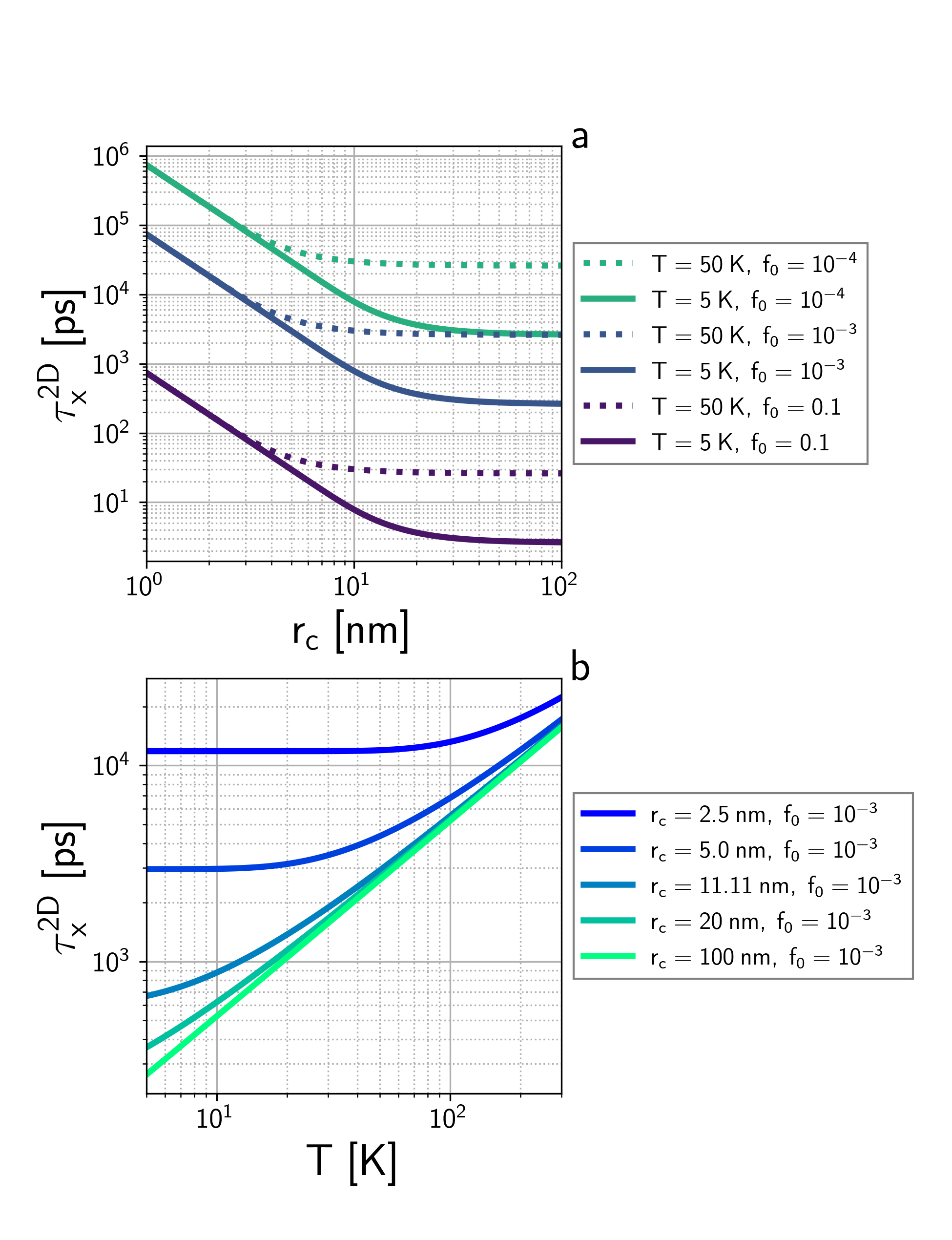}
	\caption{
		Exciton radiative lifetime vs. coherence radius (a) and temperature (b) as given by \autoref{eq:rad_dist2} for different oscillator strengths. The oscillator strengths are chosen to exemplify those expected for dark ($f_0=10^{-4}$), interlayer ($f_0=10^{-3}$), and bright ($f_0=10^{-1}$) excitons in TMDCs.    
	}\label{fig:rad_life} 
\end{figure}

All rates from excited states to the ground/gone state are assumed to be radiative. 
Dark and interlayer exciton states have much slower radiative rates than bright exciton states. 
In order to implement the fact a trapped exciton can have a longer lifetime than an untrapped exciton, we use the work of \textcite{Feldmann_Elliott_1987} which connects the idea of a \emph{coherence area}, $A_c$ to a radiative rate. 
Each radiatve rate for each exciton is calculated on the fly given the exciton's current position. 
If the exciton is in a trap, the trap radius is used to calculate the coherence area by letting
\begin{align}
A_c = \pi r_c^2 = \pi r_\text{trap}^2 \label{eq:A_c}
\end{align}
If the exciton is not in trap, its coherence radius is arbitrarily set to 100 nm; this value was chosen because it corresponds to the largest explored trap  and it is far past the saturation threshold shown in \autoref{fig:rad_life}.

The area of an exciton's coherence before scattering is\cite{Feldmann_Elliott_1987} 
\begin{align}
A_c = \frac{2\pi \hbar^2}{\Delta(T)M},
\end{align}
in which $\Delta$ is the temperature dependent homogeneous linewidth and $M=m_e^* + m_h^*$ is the exciton's reduced mass which for MoSe\textsubscript{2} is $1.25\cdot m_0$.\cite{Ramasubramaniam_2012}
The radiative decay rate for 2D free excitons is\cite{Feldmann_Elliott_1987} 
\begin{align}
\tau^{\text{2D}}_x &= \frac{\pi\epsilon_0m_0c^3}{ne^2\omega^2 f_0}\left[\frac{1}{1-\exp\left(-\frac{2\pi\hbar^2}{MA_ck_BT}\right)}\right]\left(\frac{A_x}{A_c}\right), \label{eq:rad_dist}
\end{align}
such that the effective area of the exciton is
\begin{align}
A_x = \pi\left(a_0^{\text{2D}}\right)^2 = \frac{\hbar^2}{2\mu E_B^{\text{2D}}}, \label{eq:A_x}
\end{align}
in which $\omega$ is the frequency of the radiative transition, $n$ is the refractive index into which the photon is emitted, $f_0$ is the dipole matrix element connecting Bloch states in the valence and conduction bands (the oscillator strength), $E_B^{\text{2D}}$ is the binding energy of the exciton, and $\mu = \frac{m_e^* \times m_h^*}{m_e^* + m_h^*}$. 
Combining \autoref{eq:A_c}, \autoref{eq:rad_dist}, and \autoref{eq:A_x} yields an expression for the radiative rate
\begin{align}
\tau^{\text{2D}}_x &= \frac{\epsilon_0m_0c^3\hbar^2}{2\mu E_B^{\text{2D}} r_c^2ne^2\omega^2 f_0}\left[\frac{1}{1-\exp\left(-\frac{2\hbar^2}{M r_c^2k_BT}\right)}\right] \label{eq:rad_dist2}
\end{align}
which is exclusively a function of well-known material constants along with temperature, transition dipole, and coherence (trap) radius. 
\autoref{fig:rad_life} graphs how temperature, oscillator strength, and coherence radius all conspire to give specific radiative lifetimes. 
Observe how the radiative lifetime plateaus to a minimum for $r_c \gtrsim 10 \text{ nm}$ which implies that the radiative lifetime is not heavily dependent on trap radius for traps larger than $\sim$10 nm in radius. 
Also observe that the homogeneous linewidth, $\Delta(T)$, does not appear in \autoref{eq:rad_dist2}.

\subsection{Exciton exciton annihilation}

We model exciton exciton annihilation (EEA) as a coalescence reaction
\begin{align}
X + X \xrightarrow{\text{EEA}} X.
\end{align}
This reaction occurs with a probability $P_{\text{anhil}}$ whenever two excitons are within some annihilation capture radius $r_{\text{anhil}}$.
If the reaction occurs, both excitons are moved to their joint Euclidean mid-point and one of their identities is set to ``gone''. 

There are many ways to computationally implement the idea of EEA---many of these methods have subtle differences in their outcomes.
We choose to allow every particle to only undergo EEA once per time step. 

$\mathfrak{p}_{\text{anhil}}$ starts by calculating the Euclidean distancing between all elements of $\mathbb{S}$ of the same type which yields an $N\times N$ matrix, $\mathbf{R}$.
$\mathbf{R}$ is then forced to be strictly upper triangular by setting all elements including and below the diagonal to be zero.
While keeping track of original row and column index, the elements of $\mathbf{R}$ are sorted from smallest to largest distance to create the tuple $\mathbb{R}$ with only elements smaller than $r_{\text{anhil}}$ being kept.
A set, $\mathbb{A}$, is instantiated to contain the index of each row/column of the elements of $\mathbf{R}$ which have participated in an annihilation event for the particular timestep.

Starting with the smallest element of $\mathbb{R}$, a random number, $\tilde{P}_{\text{choice}}$ drawn from a uniform distribution $[0,1)$ is calculated. 
If $P_{\text{anhil}}>\tilde{P}_{\text{choice}}$ and $\text{row, col} \notin \mathbb{A}$ then the particle numbers corresponding to the row and column undergo EEA and the row and column indices are then added to $\mathbb{A}$. 
This process is repeated for all remaining elements of $\mathbb{R}$.

\subsection{Dipole-dipole interactions}

An interlayer exciton has an electron in one layer and a hole in another; this charge separation creates a static dipole.
If multiple interlayer excitons are present, they will repulsively interact with each other through a dipole-dipole interaction.
The pair-potential, $U_{\text{dd}}$, between two parallel and equal dipole moments, $p$, separated by a distance $r$ is
\begin{align}
U_{\text{dd}}(r) =- \frac{p^2}{\epsilon\epsilon_0r^3}. \label{eq:dipolepotential}
\end{align}
The force associated with a single dipole-dipole interaction is 
\begin{align}
\vec{F} &= -\nabla U_{\text{dd}}(r) \hat{r} \\
&= - \frac{3p^2\hat{r}}{\epsilon\epsilon_0r^4}. \label{eq:dipoleforce}
\end{align}
We take a mean-field approach and assume that for a small timestep, $\delta t$, all forces are constant. 
Newton's Second law of motion, $\vec{F} = m\vec{a}$, indicates that all accelerations will also be constant.
Using standard kinematic equations of motion, under a constant acceleration with no initial velocity, the change in position of a particle is 
\begin{align}
\vec{\Delta r} &= \frac{\vec{F}(\delta t)^2}{2m}, \label{eq:dipolemove}
\end{align} 
in which $m$ is the exciton's mass.
If there are only two dipoles interacting, direct substitution of \autoref{eq:dipoleforce} into \autoref{eq:dipolemove} yields the change in position of the particles. 
However, in the case of multiple dipole-dipole interactions, the total force,
\begin{align}
\vec{F}_{i,\text{total}}=\sum_{j\neq i}\vec{F}(r_{ij}), 
\end{align} must be calculated over all particle pairs and then substituted into \autoref{eq:dipolemove}; this must be done for each particle, $i$.

If there were no traps, no more additional mechanics would be required to construct a valid $\mathfrak{p}_{\text{dipole}}$.
For each time step we can calculate a new position of all interlayer excitons and move them to those positions. 
However, this scheme does not account for the fact that dipole interactions change the potential energy of the pairs which can help an interlayer exciton leave a trap. 
We incorporate the change in potential caused by a particle moving a distance $\Delta r$ as caused by the dipole force
\begin{align}
\Delta U = \int_{r}^{r+\Delta r} \vec{F} \text{d}\vec{r}
\end{align}
into the Miller-Abrahams hopping rate which controls trapping, \autoref{eq:hoprate}. Note that because $r$ predominantly appears in the expression for $F$, $\Delta U \neq F\Delta r$.

To account for multiple particles all interacting with each other, we use a mean-field approach and assume that each $\vec{F}_{i,\text{total}}$ is constant as all particles move under repulsion and that $\vec{F}_{i,\text{total}}$ originates through interaction of $i$ with a single effective point particle a distance $r_{\text{eff}}$ away
\begin{align}
\Delta U_i &= \int_{r_{\text{eff}}}^{r_{\text{eff}}+\Delta r} \vec{F}_{i,\text{total}} \text{d}\vec{r}, \\
\vec{F}_{i,\text{total}} &= - \frac{3p^2\hat{r}_{\text{eff}}}{\epsilon\epsilon_0r_{\text{eff}}^4}, \label{eq:F_reff}\\
\Longrightarrow \Delta U_i &= \frac{p^2}{\epsilon\epsilon_0}\left[\frac{1}{\left(r_{\text{eff}}+\Delta r\right)^3}-\frac{1}{r_{\text{eff}}^3}\right] \label{eq:dU_reff}
\end{align}
In order to remove the dependence of \autoref{eq:dU_reff} on $r_{\text{eff}}$, we rearrange \autoref{eq:F_reff}  to $r_{\text{eff}} = \left(\frac{-3p^2}{F_{i,\text{total}}\epsilon\epsilon_0}\right)^{\frac{1}{4}}$, and also use \autoref{eq:dipolemove}
\begin{align}
\Delta U_i &= \frac{p^2}{\epsilon\epsilon_0}\left[\left(\left(\frac{-3p^2}{F_{i,\text{total}}\epsilon\epsilon_0}\right)^{\frac{1}{4}}+\frac{F_{i,\text{total}}(\delta t)^2}{2m}\right)^{-3}-\left(\frac{-3p^2}{F_{i,\text{total}}\epsilon\epsilon_0}\right)^{\frac{-3}{4}}\right]
\end{align}

$\mathfrak{p}_{\text{dipole}}$ therefore consists of: 
\begin{enumerate}
	\item calculating all inter-exciton Euclidian separations and angles, 
	\item calculating a total force felt by each interlayer exciton, 
	\item using that total force to calculate a change in potential energy were the particle to move the distance $\Delta r$ caused by repulsion to a new $(x,y)$ coordinate, 
	\item finally \autoref{eq:hoprate} is used to calculate which interlayer exciton are stochastically moved by the dipole-dipole interaction.
\end{enumerate}

\subsection{Parameters used in model}

\begin{table}[!htbp]
	\caption{Model parameters.}
	\label{tab:params}
	\resizebox{\textwidth}{!}{%
		\begin{tabular}{|l||l|l|l|}
			\hline
			name                                         & symbol                   & value                                          & reference(s) and notes                                                                                           \\ \hline\hline
			electron mass                                & $m_e$                    & $0.7m_0$                                       & \cite{Ramasubramaniam_2012}                                                                                      \\ \hline
			hole mass                                    & $m_h$                    & $0.55m_0$                                      & \cite{Ramasubramaniam_2012}                                                                                      \\ \hline
			exciton mass                                 & $M$                      & $m_e+m_h$                                      &                                                                                                                  \\ \hline
			radiative emission frequency                 & $\hbar\omega$            & 1500 meV                                       & Chosen to be approximately between emission of MoSe\textsubscript{2}, WSe\textsubscript{2} and their IX.\footnote{Because oscillator strengths are estimated, the exact value is inconsequential.}      \\ \hline
			refractive index                             & $n$                      & 5                                              & Estimated from \cite{Gu_Liu_2019}.                                                                               \\ \hline
			IX dipole                                    & $p$                      & 0.65 e nm                                      & \cite{Karni_Heinz_2019}                                                                                          \\ \hline
			TMDC permittivity                            & $\epsilon_{\text{TMDC}}$ & 7.2                                            & \cite{Kim_Tutuc_2015}                                                                                            \\ \hline
			exciton diffusion constant                   & $D$                      & 50 nm\textsuperscript{2}ps\textsuperscript{-1} & Informed by \cite{Zipfel_Chernikov_2020, Yuan_Huang_2017, Mouri_Matsuda_2014}. Measured $D$ vary widely in the literature.           \\ \hline
			bright-dark splitting (WSe\textsubscript{2}) & $E_{BD}$                 & - 30 meV                                       & \cite{Zhang_Heinz_2015}                                                                                          \\ \hline
			bright-IX splitting (WSe\textsubscript{2})   & $E_{BIX}$                & -300 meV                                        & Estimated from \cite{Ovesen_Malic_2019}.                                                                         \\ \hline
			simulation timestep                          & $\delta t$               & 0.05 ps                                        & Arbitrarily chosen such that it shorter than the system coherence time.\cite{Moody_Li_2015, Martin_Cundiff_2020} \\ \hline
			EEA annhiliation radius                      & $r_{\text{anhil}}$       & 1.5 nm                                         & \cite{Mouri_Matsuda_2014} and also estimated from \cite{Yuan_Huang_2017}.                                        \\ \hline
			annhiliation probability                     & $P_{\text{anhil}}$       & 0.25                                           & \cite{Mouri_Matsuda_2014}                                                                                        \\ \hline
			rate of bright to dark conversion            & $k_{BD}$                 & 1/10 ps\textsuperscript{-1}                    & \cite{Zhang_Heinz_2015}                                                                                          \\ \hline
			rate of bright to IX conversion              & $k_{BIX}$                & 1/2 ps\textsuperscript{-1}                     & Estimated from \cite{Ovesen_Malic_2019, Merkl_Huber_2019, Hong_Wang_2014}.                                       \\ \hline
			bright oscillator strength                   & $f_{0,B}$                & 0.1                                            & Estimated from \cite{Martin_Cundiff_2020, Robert_Marie_2016}.                                                    \\ \hline
			IX oscillator strength                       & $f_{0,IX}$               & 0.001                                          & Informed by \cite{Zhu_Zhu_2017, Yuan_Huang_2020, Miller_Wurstbauer_2017}.                                        \\ \hline
			dark oscillator strength                     & $f_{0,D}$                & 0.0001                                         & Estimated from \cite{Zhang_Heinz_2015}.                                                                          \\ \hline
		\end{tabular}%
	}
\end{table}

\subsection{Putting it all together: temporal evolution}

Combining the results of the previous subsections results in a set of nested propagators.
A diffraction limited Gaussian pulse with a full-width-half-maximum of 215 nm excites 1000 excitons into the bright state.
This excitation condition corresponds to an initial density of 0.02 excitons per nm\textsuperscript{2} (2$\times$10\textsuperscript{12} excitons per cm\textsuperscript{2}).
These bright excitons then convert to dark and IX states which all eventually transition to the ground (gone) state. 
As stated in the main text, in order to quantify trapping efficiency, we define a metric, 
\begin{align}
\text{trapping efficiency}_{\text{IX}} \equiv \eta_{\text{IX}} \equiv  \frac{\int{N_{\text{trapped}}\text{d}t}}{\int{N_{\text{total}}\text{d}t}}, \label{eq:trapratio}
\end{align}
which requires us to sum over the full temporal evolution of the excitons. 

\clearpage
\section{Investigation of trap shape}

\begin{figure}[!htbp]
	\centering
	\includegraphics[trim=0 1cm 0 1.5cm, clip, width=\linewidth]{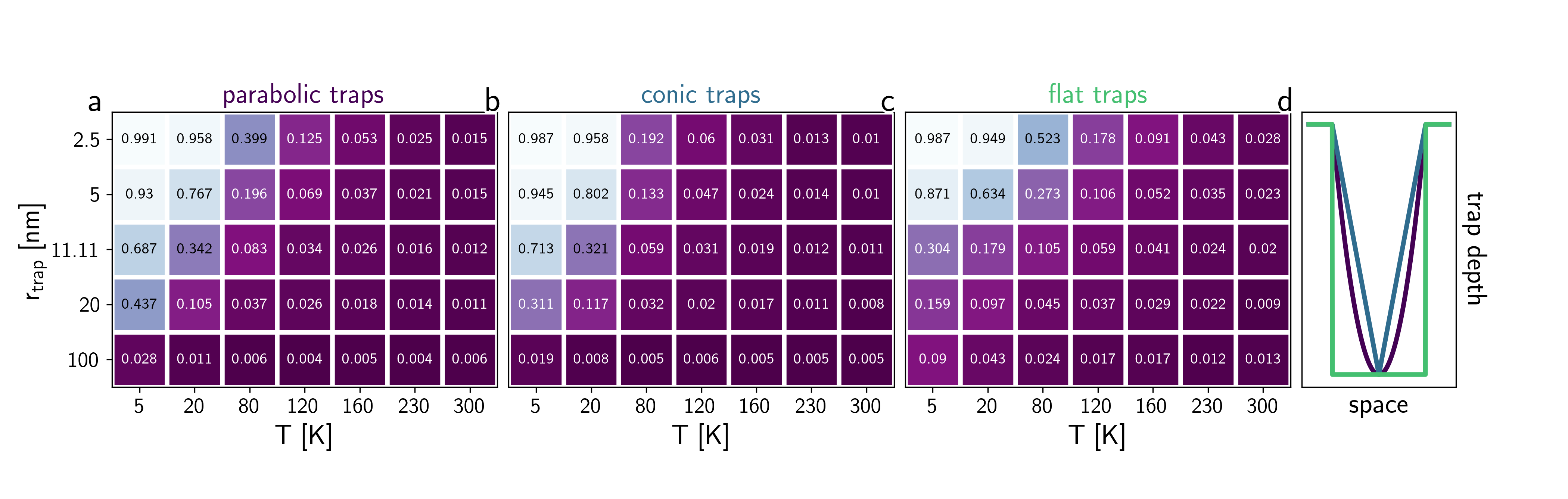}
	\caption{
		Effects of trap geometry and temperature on \textbf{interlayer exciton} trapping. 
		For all simulations, traps are 50 meV deep and the total trap area is kept constant at $\pi (100 \text{ nm})^2$.
	}\label{fig:shapedtraps} 
\end{figure}

\clearpage
\section{Trapping of dark excitons}

In order to compare the effects of traps on the long-lived dark excitons as compared to the long-lived interlayer excitons (which also have a dipole interaction), in this section we show a set of back-to-back trapping efficiency figures for dark and interlayer excitions which span our simulation parameter space.

\begin{figure}[!htbp]
	\centering
	\includegraphics[trim=0 1cm 0 2cm, clip, width=\linewidth]{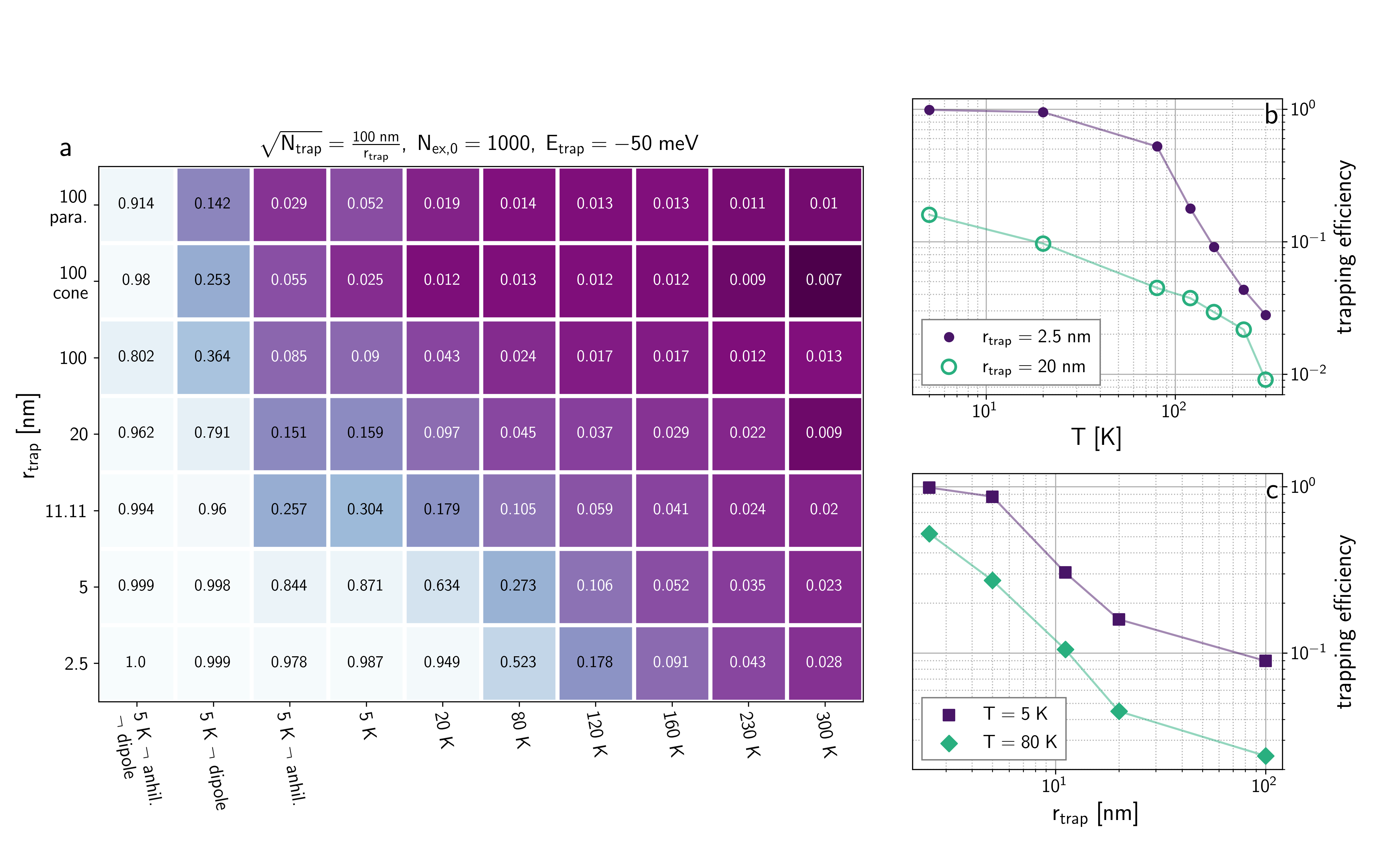}
	\caption{
		Effects of trap geometry and temperature on \textbf{interlayer exciton} trapping. 
		(a) Trapping efficiency as a function of trap radius, trap type, temperature, and exciton-exciton interactions. The total area of traps is constant for all simulations. Smaller trapping ratios map to darker colors. 
		(b) Trapping efficiency vs. temperature for 2.5 and 20 nm radius traps on a log-log scale.
		(c) Trapping efficiency vs. trap radius at 5 and 80 K on a log-log scale.
	}\label{fig:radius_v_temp_IX} 
\end{figure}

\begin{figure}[!htbp]
	\centering
	\includegraphics[trim=0 1cm 0 2cm, clip, width=\linewidth]{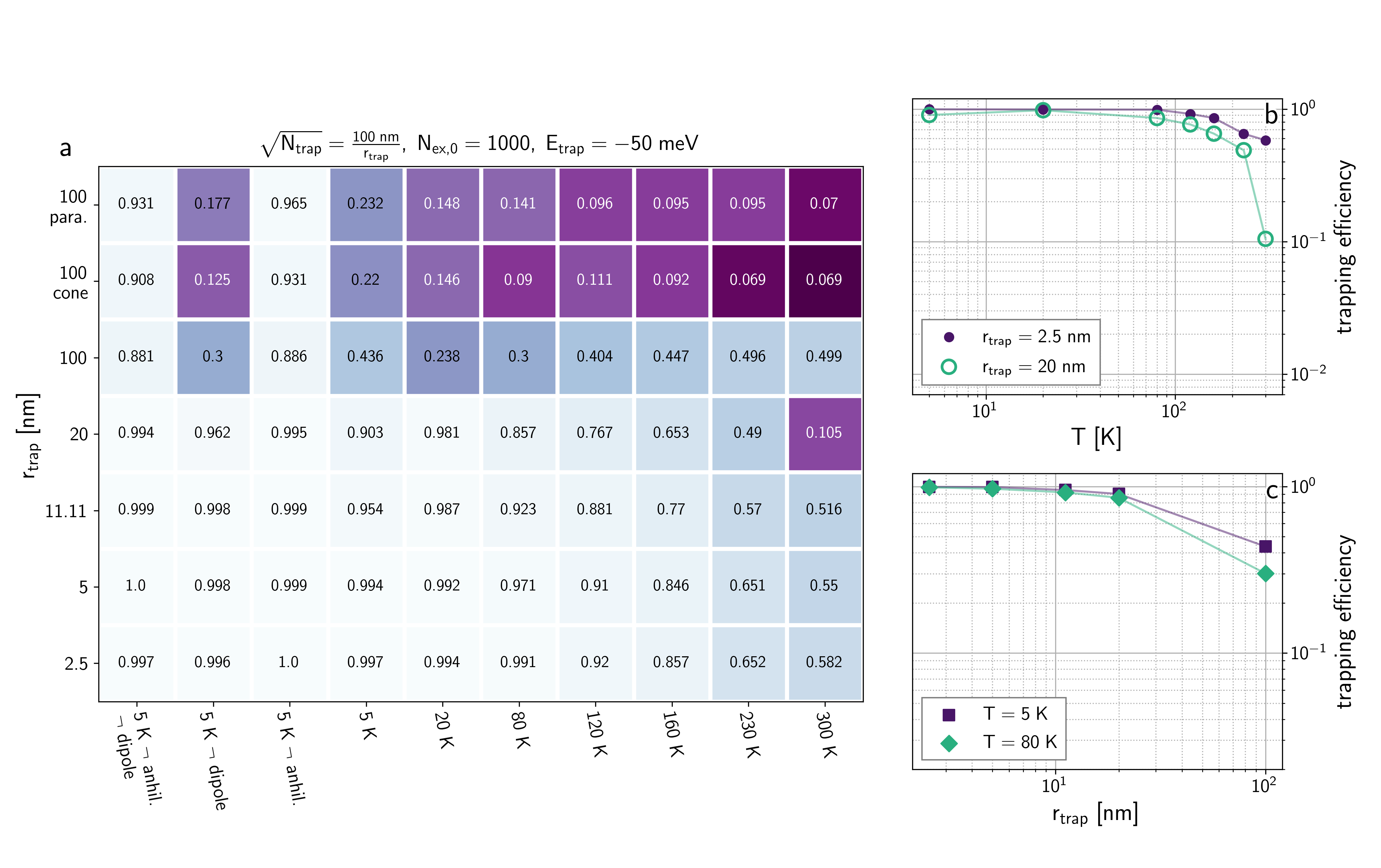}
	\caption{
		Effects of trap geometry and temperature on \textbf{dark exciton} trapping. 
		(a) Trapping efficiency as a function of trap radius, trap type, temperature, and exciton-exciton interactions. The total area of traps is constant for all simulations. Smaller trapping ratios map to darker colors. 
		(b) Trapping efficiency vs. temperature for 2.5 and 20 nm radius traps on a log-log scale.
		(c) Trapping efficiency vs. trap radius at 5 and 80 K on a log-log scale.
	}\label{fig:radius_v_temp_dark} 
\end{figure}

\begin{figure}[!htbp]
	\centering
	\includegraphics[trim=0 1cm 0 1cm, clip, width=\linewidth]{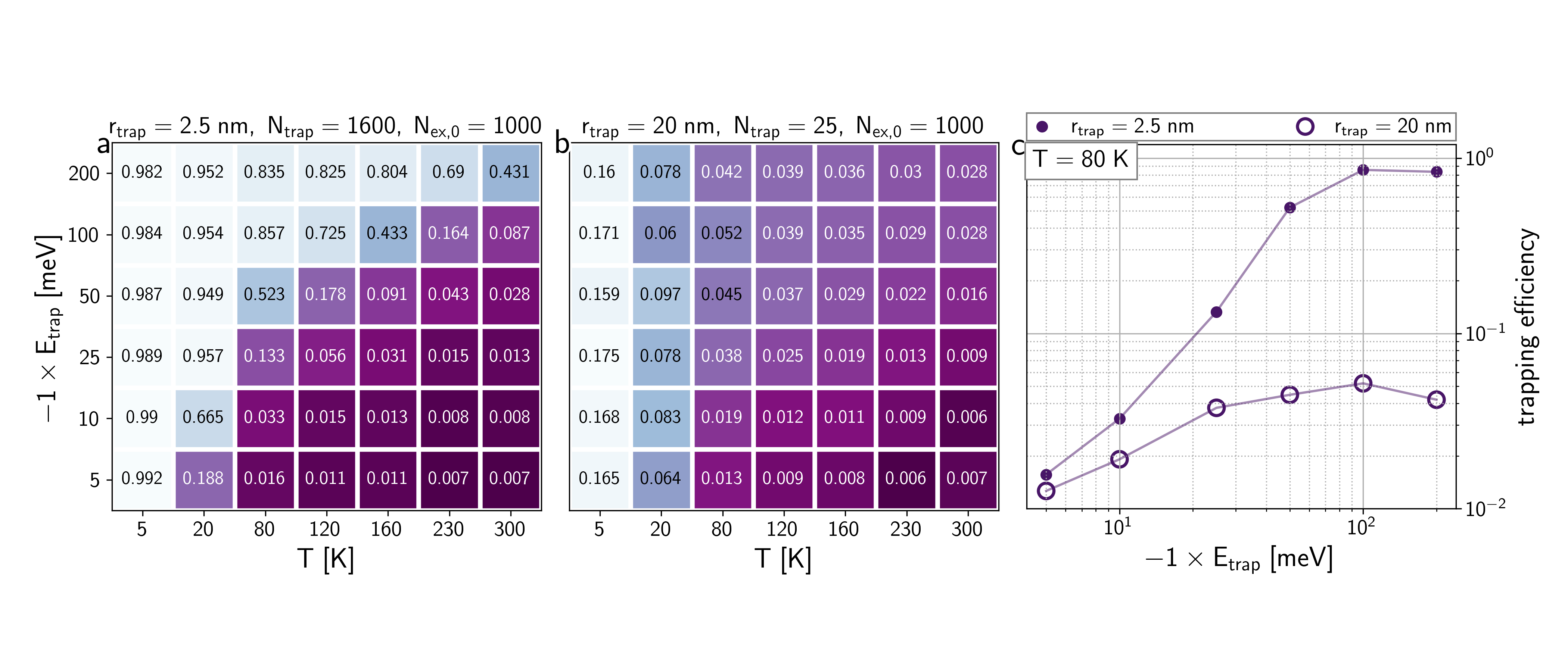}
	\caption{
		Trapping efficiency of \textbf{interlayer excitons} as a function of temperature and trap depth for 2.5 nm (a) and 20 nm (b) radius traps. Note that the numerical extent of the colormap is not shared across the sub-figures. 
		(c) Ix trapping efficiency vs. trap depth for 2.5 and 20 nm radius traps on a log-log scale at 80 K. 
	}\label{fig:depth_v_temp_IX} 
\end{figure}

\begin{figure}[!htbp]
	\centering
	\includegraphics[trim=0 1cm 0 1cm, clip, width=\linewidth]{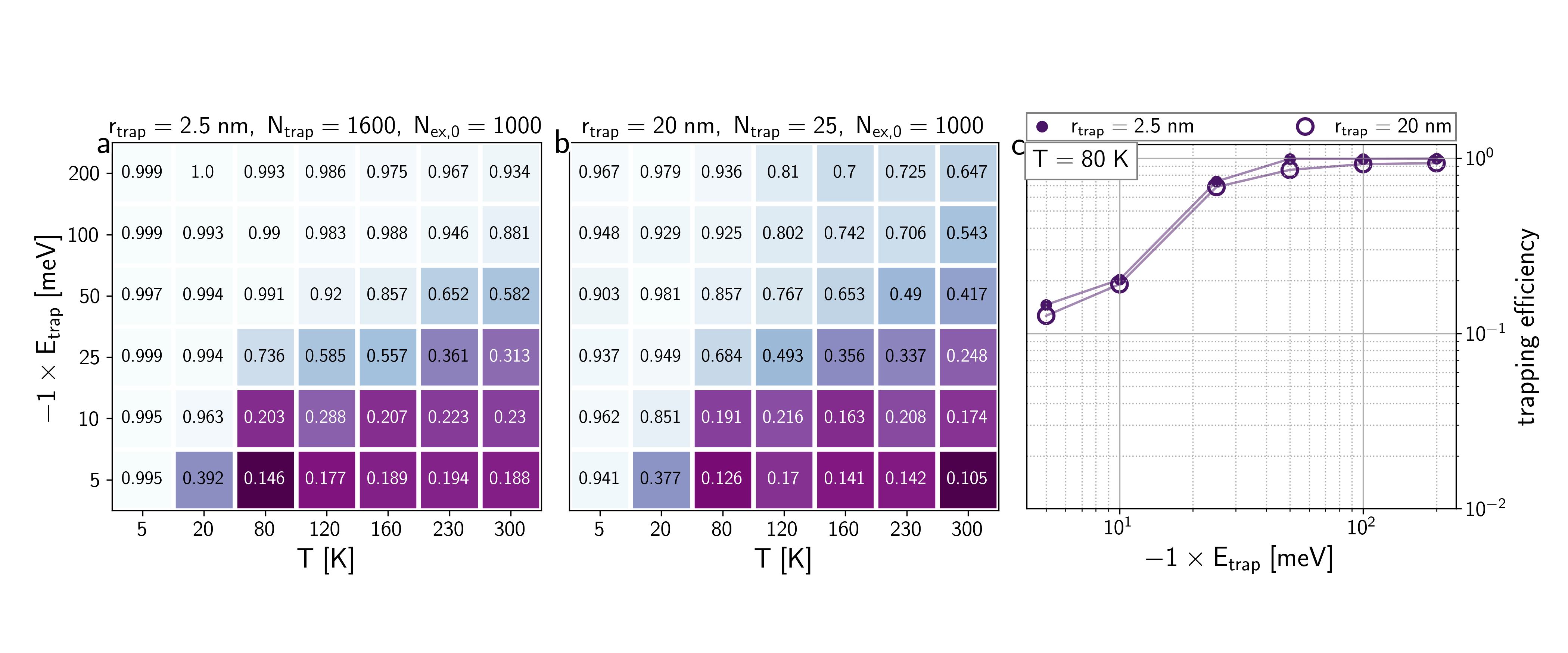}
	\caption{
		Trapping efficiency of \textbf{dark excitons} as a function of temperature and trap depth for 2.5 nm (a) and 20 nm (b) radius traps. Note that the numerical extent of the colormap is not shared across the sub-figures. 
		(c) Dark exciton trapping efficiency vs. trap depth for 2.5 and 20 nm radius traps on a log-log scale at 80 K. 
	}\label{fig:depth_v_temp_dark} 
\end{figure}

\begin{figure}[!htbp]
	\centering
	\includegraphics[trim=0 1cm 0 1cm, clip, width=\linewidth]{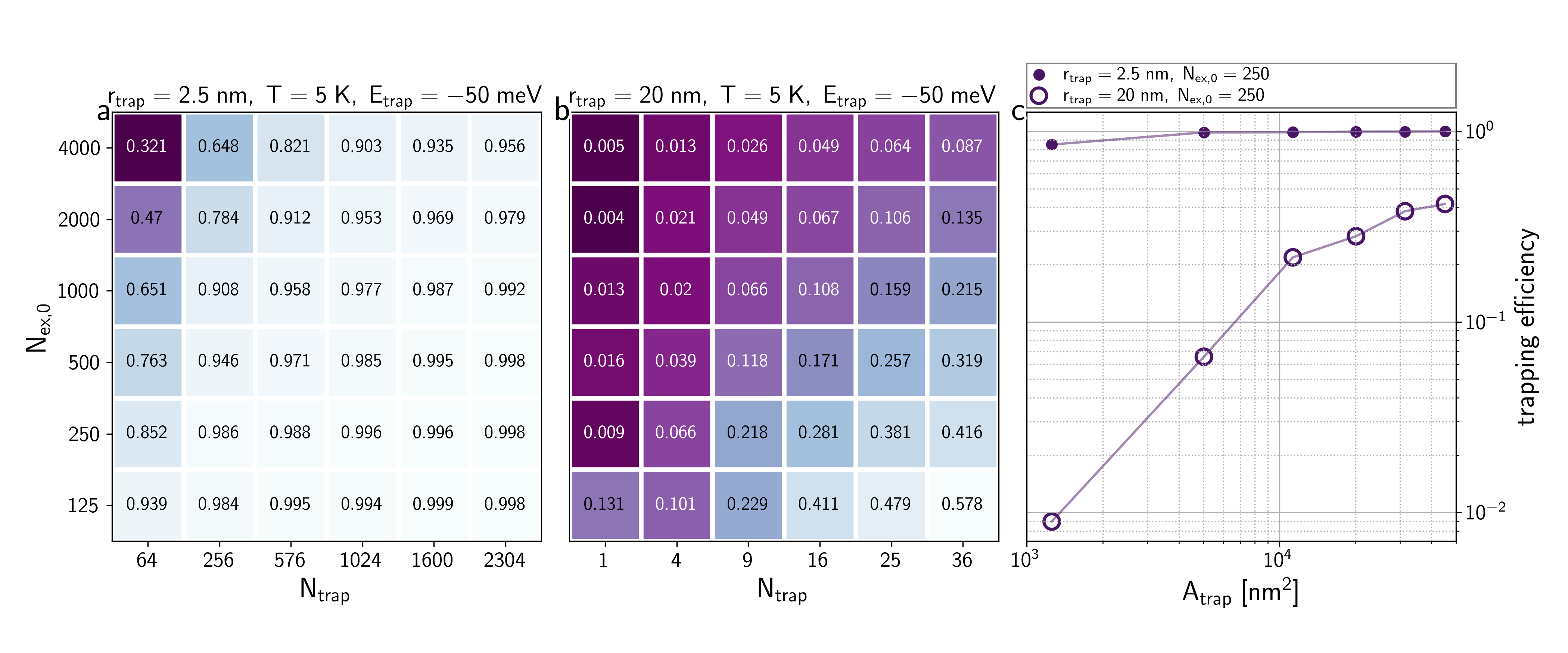}
	\caption{
		Trapping efficiency of \textbf{interlayer excitons} as a function of trap and exciton number for 2.5 nm (a) and 20 nm (b) radius traps. Note that the numerical extent of the colormap is not shared across the sub-figures, and that the ith column of (a) and (b) have the same total trap area.
		(c) Exciton trapping efficiency vs. trap area for 2.5 and 20 nm radius traps on a log-log scale for initial bright exciton populations of 250.
	}\label{fig:trapnum_v_exnum_IX} 
\end{figure}

\begin{figure}[!htbp]
	\centering
	\includegraphics[trim=0 1cm 0 1cm, clip, width=\linewidth]{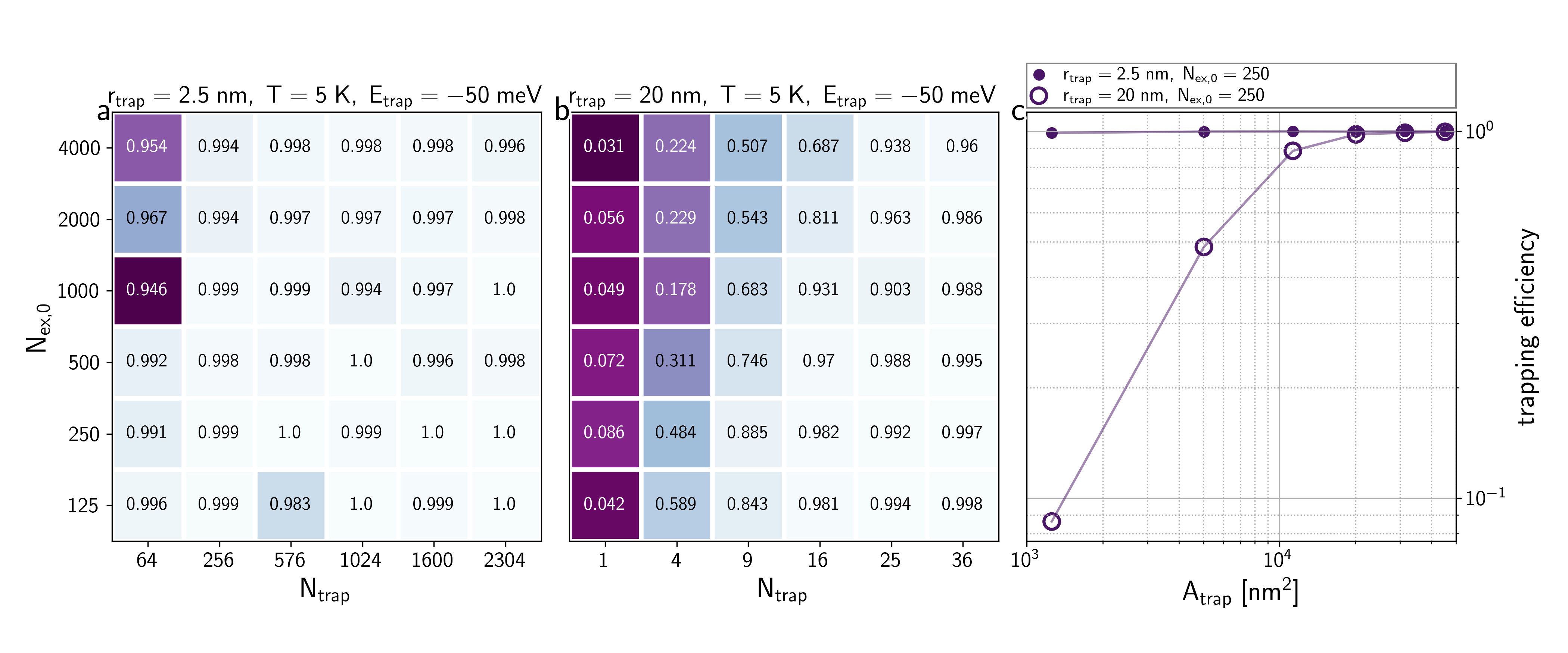}
	\caption{
		Trapping efficiency of \textbf{dark excitons} as a function of trap and exciton number for 2.5 nm (a) and 20 nm (b) radius traps. Note that the numerical extent of the colormap is not shared across the sub-figures, and that the ith column of (a) and (b) have the same total trap area.
		(c) Exciton trapping efficiency vs. trap area for 2.5 and 20 nm radius traps on a log-log scale for initial bright exciton populations of 250.
	}\label{fig:trapnum_v_exnum_dark} 
\end{figure}

\clearpage
\section{Occupancy of individual traps}

In this section we explore how trap occupancy changes as a function of time, trap radii, temperature, and trap depth.
To prepare a visualization we do the following for each simulation of interest:
\begin{enumerate}
	\item Calculate how many IXs are in each trap at each saved simulation time points. In terms of notation, the $i$th trap will have some occupancy $\Omega_i(t)$ at some time $t$.
	\item The set of traps, $\{i\}$ are sorted from low to high based on the size of $\int \Omega_i \text{d}t$. 	
\end{enumerate}

\begin{figure}[!htbp]
	\centering
	\includegraphics[trim=0 1cm 0 1cm, clip, width=\linewidth]{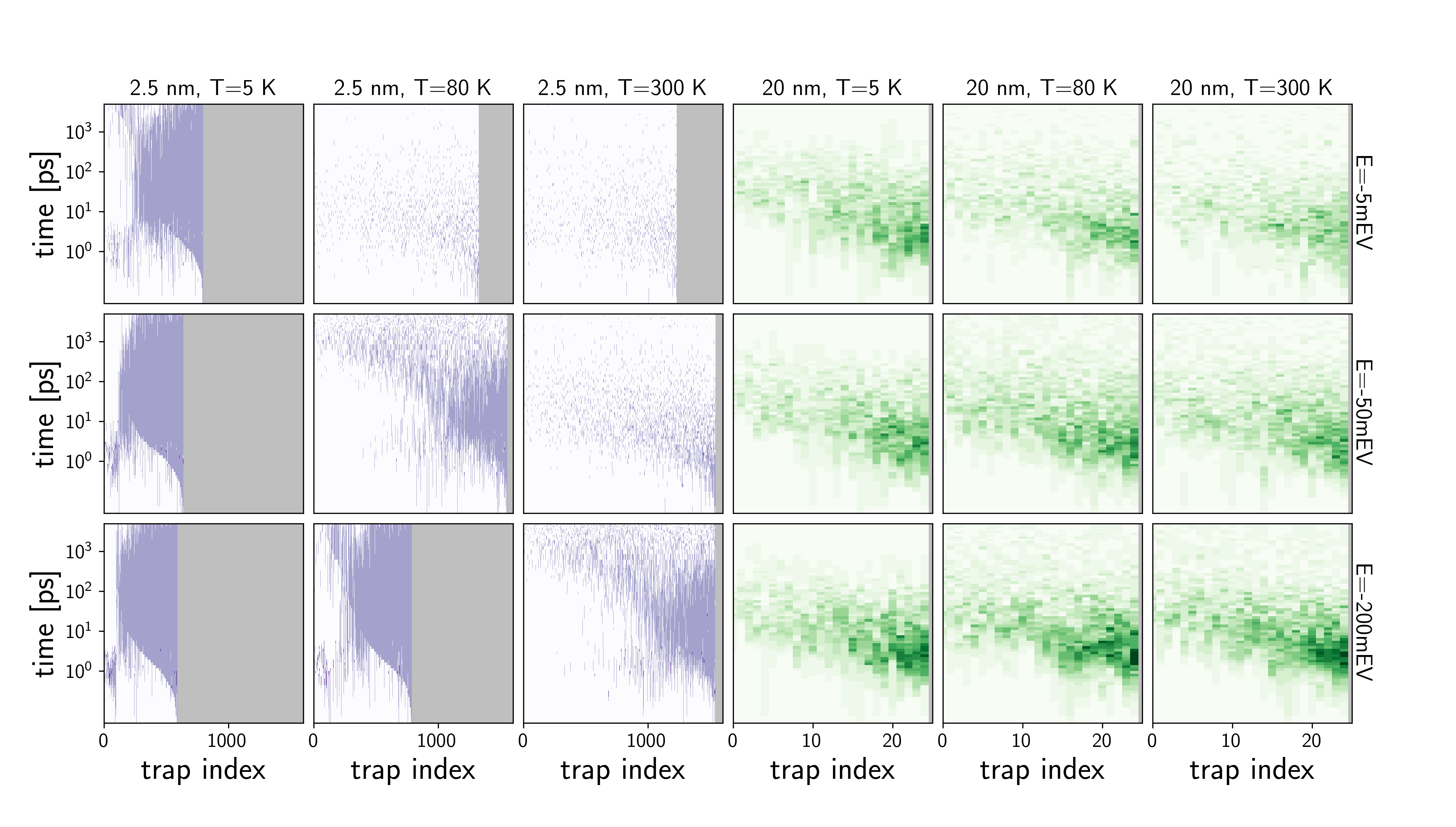}
	\caption{
		Trap occupancy vs. time. 
		Rows correspond to different trap depths.
		Columns correspond to different temperature and trap radii.
		For 2.5 nm traps (first three columns), the purple colormap spans from 0 to 2.
		For 20 nm traps (last three columns), the green colormap spans from 0 to 15. 
	}\label{fig:trap_occupancy} 
\end{figure}

\begin{figure}[!htbp]
	\centering
	\includegraphics[trim=0 1cm 0 1cm, clip, width=0.5\linewidth]{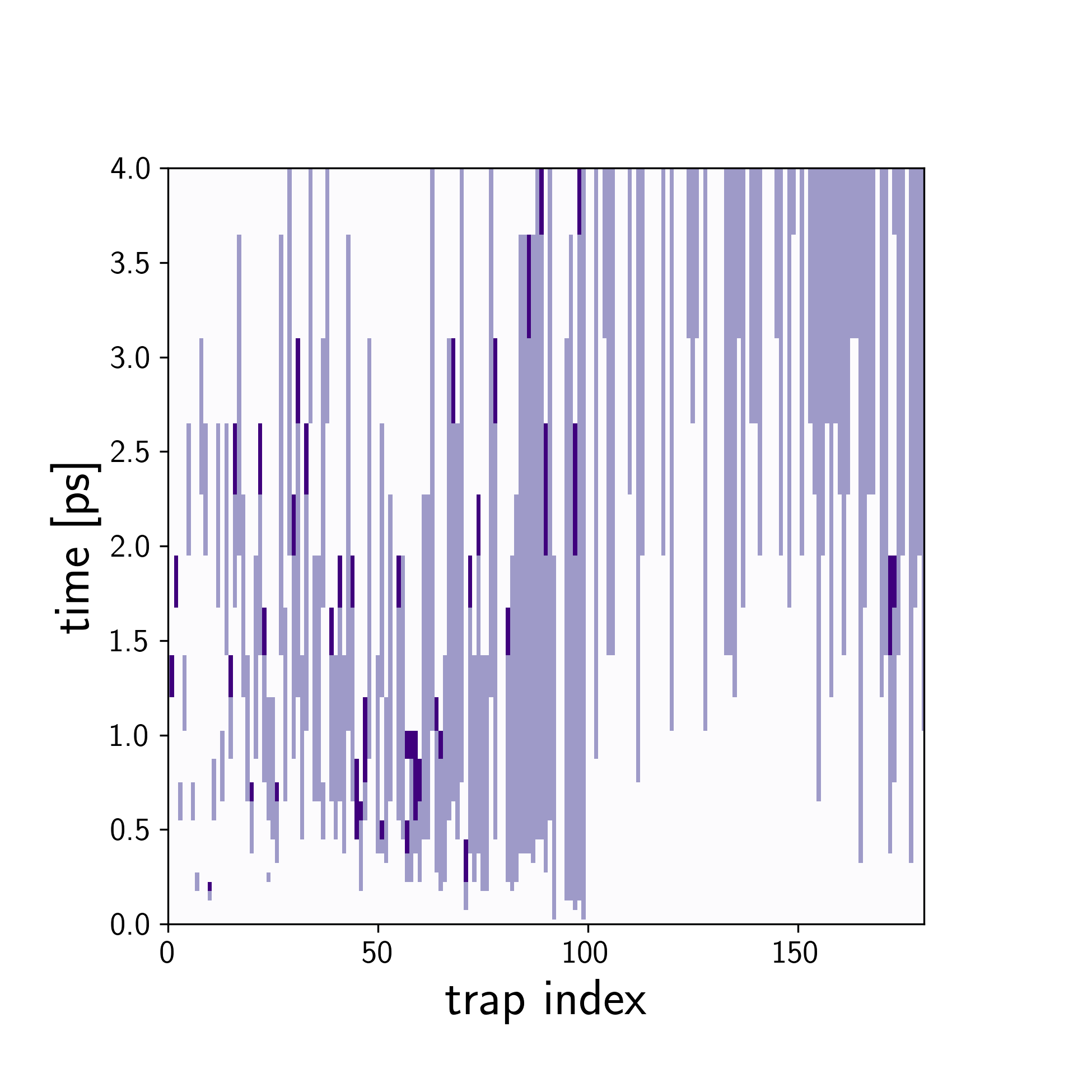}
	\caption{
		Trap occupancy vs. time. for 1600 2.5 nm, 200 meV traps at 5 K showing that double occupancy of a trap results in the subsequent time-points the trap being empty. 
		This is a zoom-in of the bottom-left dataset shown in \autoref{fig:trap_occupancy}.
	}\label{fig:trap_occupancy_zoom} 
\end{figure}

\clearpage
\bibliography{database}